\documentclass[longauth]{aa}
\usepackage[utf8]{inputenc}
\usepackage{txfonts}
\usepackage{xcolor}
\usepackage{gensymb}
\usepackage{natbib}
\usepackage{courier}
\usepackage{longtable}
\usepackage{natbib}
\usepackage[normalem]{ulem}
\def \MJ{M$_{\mathrm{Jup}}$}
\def \RJ{R$_{\mathrm{Jup}}$}
\def \RS{R$_{\odot}$}

\def \LS{L$_{\odot}$}

\def \1s{$1\,\sigma$}

\def \t0{T$_0$}

\begin{document} 

\title{TOI-1296b and TOI-1298b observed with TESS and SOPHIE\thanks{Based on observations collected with the SOPHIE spectrograph on the 1.93 m telescope at the Observatoire de Haute-Provence (CNRS), France.}: Two hot Saturn-mass exoplanets with different densities around metal-rich stars }

\author{C. Moutou$^1$, J.M. Almenara$^2$, G. H\'ebrard$^{3,26}$, N.C. Santos$^{4,22}$, K.G. Stassun$^5$, S. Deheuvels$^1$, S. Barros$^4$, P. Benni$^6$, A. Bieryla$^7$, I. Boisse$^8$, X. Bonfils$^2$, P.T. Boyd$^9$, K.A. Collins$^7$, D. Baker$^{10}$, P. Cort\'es-Zuleta$^8$, S. Dalal$^3$, F. Debras$^1$, M. Deleuil$^8$, X. Delfosse$^2$, O. Demangeon$^4$, Z. Essack$^{11,12}$, T. Forveille$^2$, E. Girardin$^{13}$, P. Guerra$^{14}$, N. Heidari$^{23,8,24}$,, K. Hesse$^{12}$, S. Hoyer$^8$, J.M. Jenkins$^{15}$, F. Kiefer$^{16}$, P. C. K\"onig$^3$, D. Laloum$^{17}$, D. Latham$^7$, T. Lopez$^8$, E. Martioli$^{3,25}$, H.P. Osborn$^{12,18}$, G. Ricker$^{12}$, S. Seager$^{12}$, R. Vanderspek$^{12}$, M. Vezie$^{12}$, J. Villase{\~ n}or$^{12}$, J. Winn$^{19}$,  B. Wohler$^{15,20}$, C. Ziegler$^{21}$ }

\institute{
\inst{1} Universit\'e de Toulouse, CNRS, IRAP, 14 avenue Belin, 31400 Toulouse, France, \email{claire.moutou@irap.omp.eu} \\
\inst{2} Universit\'e Grenoble Alpes, CNRS, IPAG, 38000 Grenoble, France\\
\inst{3}  Institut d’Astrophysique de Paris, CNRS, UMR 7095, Sorbonne
Universit\'e, 98 bis bd Arago, 75014 Paris, France \\
\inst{4} Instituto de Astrof\'isica e Ci\^encias do Espa\c{c}o, Universidade do Porto, CAUP, Rua das Estrelas, 4150-762 Porto, Portugal\\
\inst{5} Vanderbilt University, Department of Physics \& Astronomy, 6301 Stevenson Center Ln., Nashville, TN 37235, USA\\
\inst{6} Acton  Sky  Portal  (Private  Observatory), Acton, MA, USA\\
\inst{7} Center for Astrophysics, Harvard \& Smithsonian, 60 Garden Street, Cambridge, MA 02138, USA\\
\inst{8} Aix-Marseille   Universit\'e,   CNRS,   CNES,   LAM   (Laboratoire d’Astrophysique de Marseille),  Marseille, France \\
\inst{9} Astrophysics Science Division, NASA Goddard Space Flight Center, Greenbelt, MD 20771, USA\\
\inst{10} Physics Department, Austin College, Sherman, TX 75090, USA\\
\inst{11} Department of Earth, Atmospheric and Planetary Sciences, Massachusetts Institute of Technology, Cambridge, MA 02139, USA\\
\inst{12} Kavli Institute for Astrophysics and Space Research, Massachusetts Institute of Technology, Cambridge, MA 02139, USA\\
\inst{13} Grand Pra Observatory, 1984 Les Hauderes, Switzerland\\
\inst{14} Observatori Astron\`omic Albany\`a, Cam\`i de Bassegoda S/N, Albany\`a 17733, Girona, Spain\\
\inst{15} NASA Ames Research Center, Moffett Field, CA, 94035, USA\\
\inst{16} LESIA, Observatoire de Paris, Universit\'e PSL, CNRS, Sorbonne Universit\'e, Universit\'e de Paris, Meudon, France\\
\inst{17} SAF, Observatoire Priv\'e du Mont, 40280, Saint-Pierre-du-Mont, France\\
\inst{18} NCCR/PlanetS, Centre for Space \& Habitability, University of Bern, Bern, Switzerland\\
\inst{19} Department of Astrophysical Sciences, Princeton University, 4 Ivy Lane, Princeton, NJ 08544, USA\\
\inst{20} SETI Institute, Mountain View, CA 94043, USA\\
\inst{21} Department of Physics, Engineering and Astronomy, Stephen F. Austin State University, 1936 North St, Nacogdoches, TX 75962, USA\\
\inst{22} Departamento de F\'isica e Astronomia, Faculdade de Ci\^encias, Universidade do Porto, Rua do Campo Alegre, 4169-007 Porto, Portugal\\
\inst{23} Department of Physics, Shahid Beheshti University, Tehran, Iran. \\
\inst{24} Laboratoire J.-L. Lagrange, Observatoire de la C\^ote d’Azur (OCA), Universite de Nice-Sophia Antipolis (UNS), CNRS, Campus Valrose, 06108 Nice Cedex 2, France\\
\inst{25} Laborat\'{o}rio Nacional de Astrof\'{i}sica, Rua Estados Unidos 154, 37504-364, Itajub\'{a} - MG, Brazil\\
\inst{26} Observatoire de Haute Provence, 04870 Saint Michel l'Observatoire, France
}
\date{April 2021}

\abstract
{We present the discovery of two new transiting extrasolar planet candidates identified as TOI-1296.01 and TOI-1298.01 by the Transiting Exoplanet Survey Satellite (TESS). The planetary nature of these candidates has been secured with the SOPHIE high-precision spectrograph through the measurement of the companion's mass with the radial velocity method. Both planets are similar to Saturn in mass and have similar orbital periods of a few days. They, however, show discrepant radii and therefore different densities. The radius discrepancy might be explained by the different levels of irradiation by the host stars. The subgiant star TOI-1296 hosts a low-density planet with 1.2~\RJ\ while the less luminous, lower-size star TOI-1298 hosts a much denser planet with a 0.84~\RJ\ radius, resulting in bulk densities of 0.198 and 0.743~g.cm$^{-3}$,  respectively. In addition, both stars are strongly enriched in heavy elements, having metallicities of $+$0.44 and $+$0.49 dex, respectively. The planet masses and orbital periods are $0.298\pm0.039$~\MJ\ and $3.9443715\pm5.8\;10^{-6}$ days for TOI-1296b, and $0.356\pm0.032$~\MJ\ and $4.537164\pm1.2\;10^{-5}$~days for TOI-1298b. The mass measurements have a relative precision of better than  13\%.}


\titlerunning{TOI-1296b and TOI-1298b}
\authorrunning{C. Moutou et al}

\maketitle

\section{Introduction}
While radial-velocity (RV) surveys were the first to discover exoplanets \citep{mayorqueloz1995}, ground-based and space-based all-sky surveys have been extremely efficient at providing a large sample of diverse transiting extrasolar systems since early 2000. Pioneering surveys carried out with OGLE \citep{udalski2002}, WASP \citep{cameron2007}, and CoRoT \citep{borde2003,deleuil2018} have first revealed the diversity of transiting Jupiter-like planets, and in particular the radius inflation for the most close-in planets \citep{moutou2013,millholland2020}. Then, the more sensitive Kepler mission unveiled the large population of coplanar super-Earth planets \citep{latham2011,rowe2015} and their characteristics \citep{howard2012}, such as the bimodal radius distribution \citep{fulton2017} and compact multiple systems \citep{lissauer2011}. The time has then come to explore individual systems with great precision in order to prepare the atmospheric characterisation of exoplanets with space observatories such as the James Webb Space Telescope \citep{colon2020} and the ARIEL satellite \citep{tinetti2020}. In this respect, the ongoing Transiting Exoplanet Survey Satellite (TESS) programme \citep{ricker} is well-suited to discover transiting systems around the brightest host stars in the whole sky. Each transiting planet discovered by TESS is worth many of its counterparts orbiting fainter stars, as the precision in the parameters' measurement is greater.\\

TESS reaches a precision of 20~ppm for bright targets in 1~h and the main goal of the mission is the detection of small planets that transit bright stars \citep{ricker}. While millions of stars are accessible for photometry through the full-frame image mode scanning the whole sky every 30 min, a selection of 200,000 stars is monitored in a 2-min cadence. Based on the observing strategy and stellar catalogues, a simulation of the planet yield has been refined by \citet{sullivan2015} and \citet{barclay2018}, and it shows that more than 14,000 planets are to be discovered by TESS, 1250 of which are expected in the 2-min cadence mode. At the time of writing, after TESS has been observing for its primary 2-year mission and 6 months into its extended programme, there are 2453 TESS objects of interest (TOI) that are publicly available to the community and under scrutiny for analysis, complementary observations, and system's characterisation. The TESS follow-up programme (TFOP) coordination is a wide community effort to optimise the work on TESS transiting planet candidates. The present work is done in this context.\\

While planet validation can be done in multiple ways \citep[e.g.][]{deeg2009,morton2012,santerne2015}, the characterisation of the transiting candidates by complementary RV observations is the main avenue \citep[e.g.][]{queloz2009,weiss2014}, and it has the huge advantage of providing information on the planet mass in addition to the radius measured by the transit method. Mass determination is key to inform planet formation models, dynamical evolution, and atmospheric characterisation. The mass measurement of transiting planets by the RV method is also unaffected by the usual sky-projection limitation of RV planets (the measured quantity being M$\times$sin$i$), since the orbit inclination angle $i$ is precisely constrained by the transit measurement. The TESS mission objective of focussing on bright stars greatly improves the ability of RV complementary observations, even with modest-size telescopes, compared to the much fainter Kepler candidates - most of which are out of reach for precise RV mass measurements. \\

In this paper, we report the discovery of two Saturn-like planets of a similar size and orbiting similar host stars, with transits revealed by TESS and RV signatures subsequently measured by the SOPHIE spectrograph. In Section 2, we present the observational material. In Section 3, the planet model and parameters are described. In Section 4, we put these two planets in context and conclude.

\begin{figure*}
\centering
\includegraphics[width=0.9\hsize]{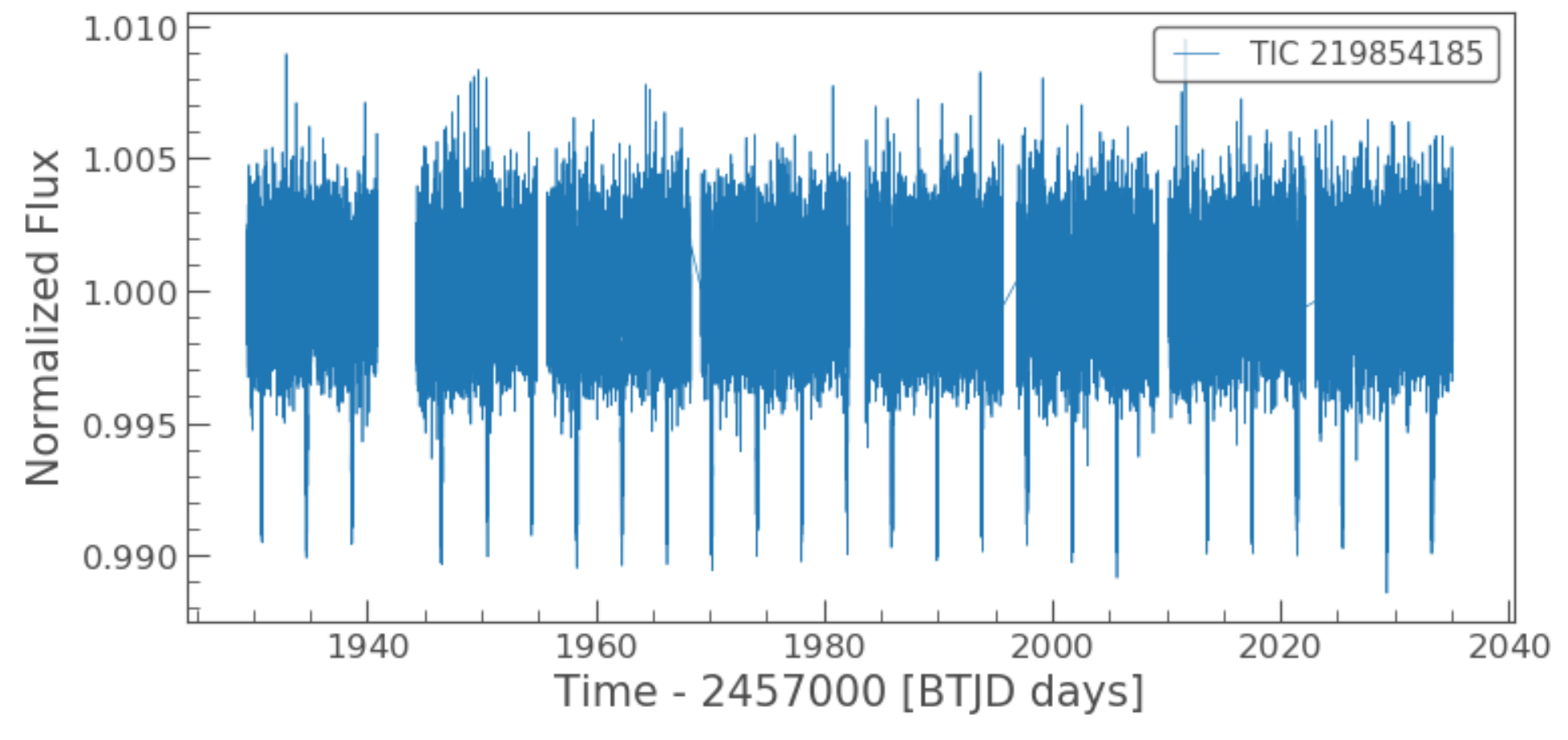}
\includegraphics[width=0.9\hsize]{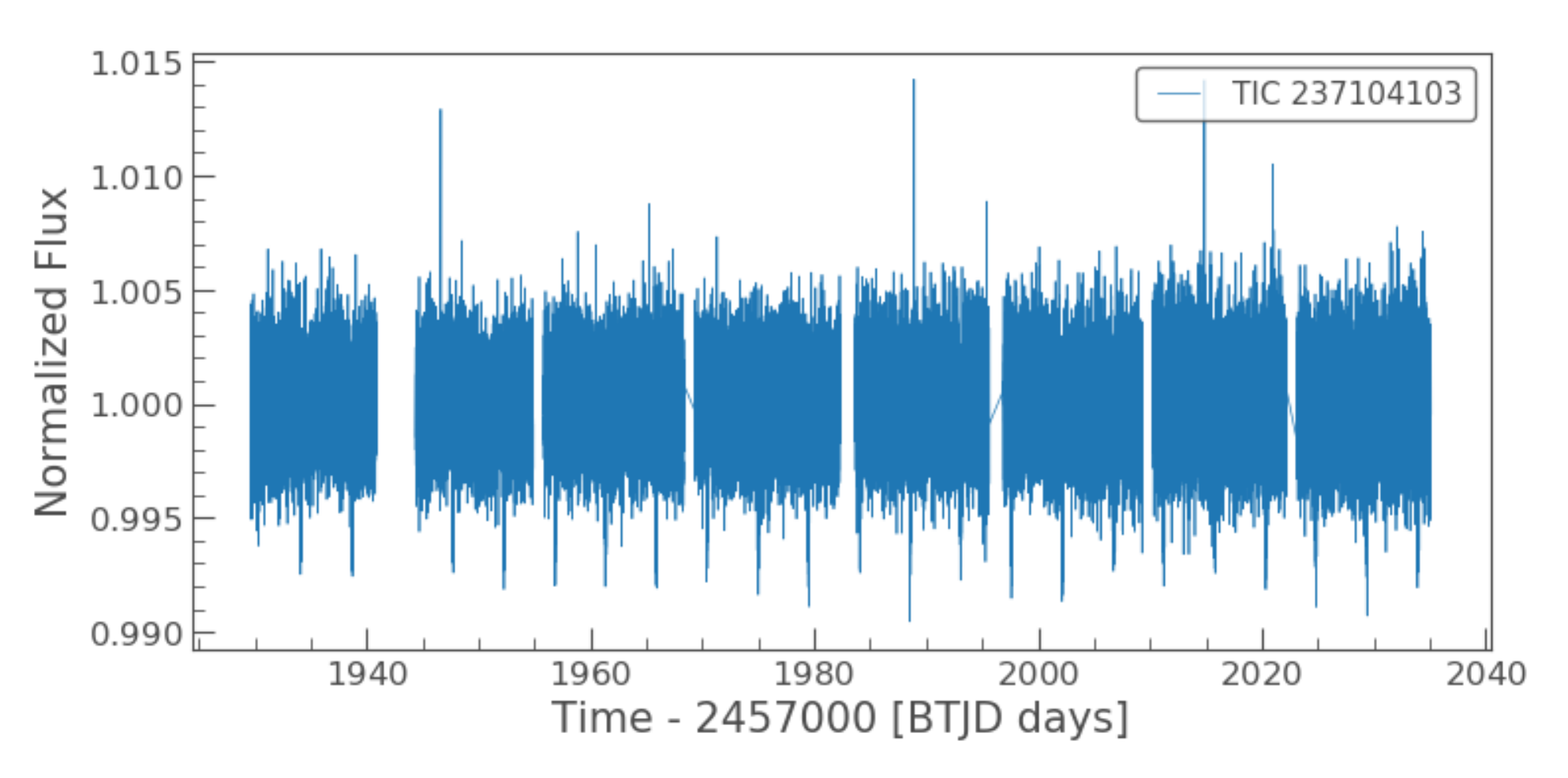}
\caption{PDC-SAP light curve of TOI-1296.01 (top) and TOI-1298.01 (bottom) obtained at a 2-min cadence.}
\label{TOI1298-TESS}
\end{figure*}

\begin{table}
\small
\begin{center}{
\caption{Properties of TOI-1296 and TOI-1298 stars.\ The values derived from this study are below the horizontal
line.\label{tab:stars}
}
\begin{tabular}{lcc}
 & TOI-1296 &  TOI-1298 \\
\hline
Gaia  & 1650851223740149504  & 1647136729864424576 \\
2MASS  &  J17070502+7014186 & J16051761+7011235 \\
TIC  & 219854185  & 237104103 \\ 
RA  & 17:07:05.018& 16:05:17.607 \\
DEC  & +70:14:18.64& +70:11:23.46\\
Distance (pc) & 313.7$\pm$1.2 & 319.0$\pm$2.0 \\ 
$B$ & 12.32$\pm$0.20 & 12.15$\pm$0.14\\
$V$ & 11.37$\pm$0.10 & 11.89$\pm$0.16\\
TESS mag & 10.8$\pm$0.01& 10.96$\pm$0.01\\
$G_{\textrm{mag}}$ &11.2912$\pm$0.0003& 11.4038$\pm$0.0003\\
$J_{2M}$ & 10.15$\pm$0.018& 10.385$\pm$0.021\\
$H_{2M}$ & 9.837$\pm$0.017& 10.105$\pm$0.017\\
$K_{2M}$ & 9.742$\pm$0.016& 10.012$\pm$0.016\\ \hline
Mass (M$_\odot$) & 1.17$\pm$0.14 & 1.44$\pm$0.10\\
Radius (R$_\odot$) &1.664$\pm$0.041 & 1.412$\pm$0.033 \\
log $g$ & 4.05$\pm$0.10& 4.39$\pm$0.08\\
$[$Fe/H$]$  & 0.44$\pm$0.04& 0.49$\pm$0.03 \\
$v$sin$i$ & 4.0$\pm$0.05  &  4.6$\pm$0.05    \\
$v_{tur}$& 1.224$\pm$0.059&1.165$\pm$0.048 \\
T$_{\rm eff}$ (K) & 5603$\pm$47& 5889$\pm$43 \\
Lbol  (L$_\odot$) & 2.455$\pm$0.088 & 2.156$\pm$0.077 \\
Age (Gy) & 6.9$\pm$0.7 & 1.6$\pm$0.9 \\ \hline
\end{tabular}
}\end{center}
\end{table}

\section{Observations}
\subsection{TESS photometry}
TOI-1298 and TOI-1296 are both stars from the selected candidate target list, and they have been observed at a 30-minute cadence between mid-July 2019 and March 2020. Periodic transit events have been quickly identified and vetted by the TESS project. When the transit events had been detected, the high cadence mode was triggered and both stars were then observed at a 2-minute cadence over a time span of 106 days in sectors 23 to 26, from March 20 to July 4, 2020. For both systems, one transit was lost in an interruption of the time series during sector 26. TOI-1296b and TOI-1298b show 25 and 23 transits in total, respectively. 
The TESS light curves for both stars at 2-minute cadence are shown in Figure \ref{TOI1298-TESS}. The images of the aperture used in obtaining the precision photometry are shown in Figure \ref{imagettes}. The systematic error-corrected PDC-SAP light curves \citep{stumpe2012,stumpe2014,smith2012} were used through the Python package \texttt{lightkurve} \citep{lightkurve2018}. 
The transits show a periodicity of 3.944 days and a depth of 0.77\% for TOI-1296.01 and a period of 4.537 days and a depth of 0.42\% for TOI-1298.01.

Preliminary vetting by the TESS project team has proven both events, TOI-1296.01 and TOI-1298.01, to be bona fide transiting planet candidates. For instance, no centroid displacement was observed, and the odd and even transit events have an identical depth within the measurement errors. The SPOC pipeline \citep{jenkins2016,twicken2018,Li:DVmodelFit2019}, as reported at MAST\footnote{https://exo.mast.stsci.edu}, clearly excludes instrumental false positives and the grazing binary scenario.\\

\subsection{Ground-based photometry}
Since both transiting events are deep and have a short period, they are easily amenable to a detection from small ground-based observatories. Scheduled photometry has thus confirmed for both candidates that the transit was occurring on the main target of the TESS aperture, rejecting the background eclipsing binary astrophysical false positive scenario. 
We used the TESS Transit Finder, which is a customised version of the Tapir software package \cite[][]{Jensen:2013}, to schedule the observations. The photometric data were extracted using the AstroImage (AIJ) software package \cite[][]{Collins:2017}.

For TOI-1296b, two full transits at the expected ephemeris were observed with 40-cm telescopes and four partial transits targeting the ingress were observed with 40 and 61-cm telescopes. The first full transit was observed at Acton Sky Portal observatory using the $r'$ filter on March 22, 2020. The ephemeris and transit depth were as expected. The second full transit was observed at Grand-Pra observatory with a 0.4-m telescope in the $g'$ filter on April 6, 2020. In the modelling, we also include the partial transits obtained at Las Cumbres observatories on March 27, 2020, and March 18, 2020, and at Adams Observatory on August 11, 2020 in the $B$ and $I_c$ filters. In March 2021, new full transits have been observed. They are not included in the present analysis.

\begin{figure}
\centering
\includegraphics[width=0.9\hsize]{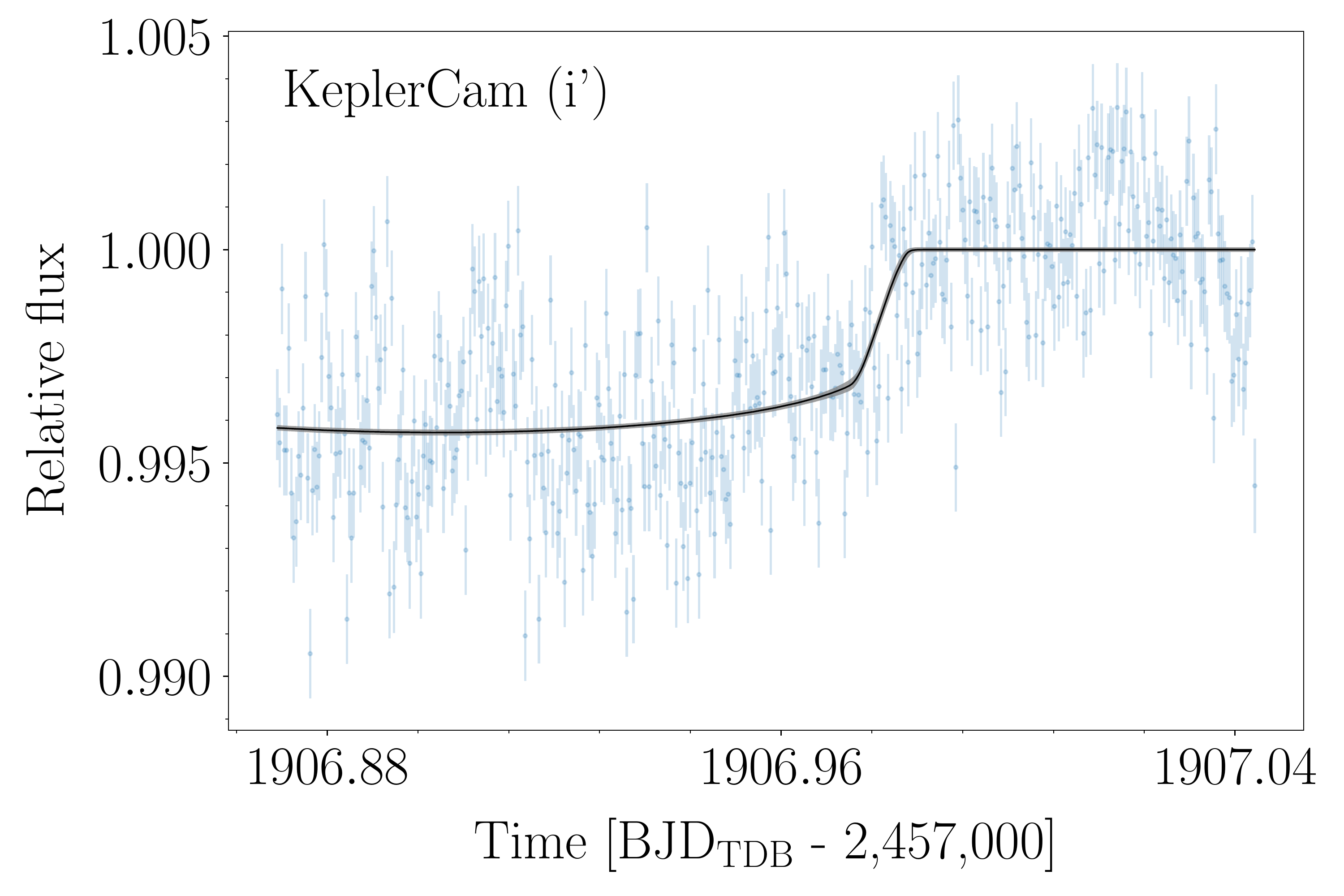}
\caption{Light curve of TOI-1298 obtained by KeplerCam, after normalisation by a straight line. The black-line model uses the posterior parameters of the system, as discussed in section 3.2. }
\label{TOI1298-keplercam}
\end{figure}

For TOI-1298b, three full transits and one partial transit were observed. The full transits are of low quality and were not used in this analysis, as they induced a bias in the results. The last ground-based transit was obtained by KeplerCam on February 27, 2020, and is shown in Figure \ref{TOI1298-keplercam}. It definitely confirms the transit to occur on the stellar host at the expected ephemeris. As it is not a full sequence, it is not included in the analysis either. 
More information on the ground-based photometric follow-up of those TOIs is available on the TFOP page\footnote{https://exofop.ipac.caltech.edu/tess/}.

\subsection{Reconnaissance spectroscopy}

We obtained reconnaissance spectra of TOI-1296 and TOI-1298 with the Tillinghast Reflector Echelle Spectrograph \citep[TRES;][]{furesz} located at the Fred Lawrence Whipple Observatory (FLWO) in Arizona, USA. TRES is a fibre-fed spectrograph with a resolving power of $\sim$44,000. The spectra were reduced and extracted as described in \citet{buchhave2010}. Relative RVs were derived by cross-correlating each observed spectrum order-by-order against the highest signal-to-noise ratio (S/N) observed spectrum. The RVs for both candidates excluded velocity variations of an amplitude compatible with a binary companion and indicated that they were good targets suitable for precision-RV follow-up observations. More specifically, two measurements of TOI-1296 were obtained with TRES with a 6 day interval, an error of 38 m/s, and a non-significant velocity difference of 17 m/s. For TOI-1298, three TRES measurements were obtained within 55 days, with an $rms$ of 32 m/s and individual errors of 25 m/s. These few data points are not included in the global fit.

The Stellar Parameter Classification \citep[SPC;][]{buchhave2012} tool was used to derive stellar parameters from the weighted average spectrum of S/N 35 and 55 for TOI-1296 and TOI-1298, respectively. SPC cross correlates an observed spectrum against a grid of synthetic spectra based on Kurucz atmospheric models \citep{kurucz1992}. The weighted average results derived with the effective temperature ($T_{\rm eff}$), metallicity ([M/H]), and surface gravity ($\log g$) and vsini as free parameters are, for TOI-1296, T$_{\rm eff}=5646\pm50$~K, log(g) = $4.19\pm0.10$, [M/H] = $0.50\pm0.08$, and vsini = $4.0\pm0.5$~km/s. For TOI-1298, the values are as follows: T$_{\rm eff}=5830\pm50$~K, log(g) = $4.29\pm0.10$, [M/H] = $0.46\pm0.08$, and vsini = $4.6\pm0.5$~km/s.

\subsection{SOPHIE spectroscopy}
High-resolution spectroscopy and RV measurements were performed at Observatoire de Haute Provence (France) for both stars with the SOPHIE (Spectrographe pour l’Observation des Ph\'enom\`enes des Int\'erieurs stellaires et des Exoplan\`etes) spectrograph \citep{perruchot2008, bouchy2013}. 
TOI-1296 was observed from June 2 to October 18, 2020, and TOI-1298 was observed from July 8 to October 26, 2020. The high-resolution mode was used for these observations, corresponding to a spectral resolution of 76,500. While the science target was observed in the main fibre, the sky background was observed in the secondary fibre. This allowed us to correct for potential contamination from the sky background, as is routinely done in this mode with fibre-fed spectrographs  \citep{bonomo2010}. With regular spectral calibrations of the spectrograph through the night using the Fabry-P\'erot etalon, the instrumental drift was also corrected at a level smaller than 1 m/s, which is much better than the photon noise RV errors for these faint stars.
Exposure times were about 1200~s and 1500~s per visit for TOI-1296 and TOI-1298, respectively. Signal-to-noise ratios of typically 20-30 per pixel were obtained at 550~nm, corresponding to photon noise uncertainties of about 3 m/s for both stars. One outlier measurement was removed from the data set for each star, corresponding to RV errors of 1.5 to 2 times the average error bars. The RVs were obtained by cross-correlating the extracted spectra with a binary mask corresponding to a G2 spectral type, which contains 3645 lines.  There are 14 RV measurements for each time series.
The bisector slope was also measured on the cross-correlation functions \citep{queloz2001}. The SOPHIE RVs are listed in Tables \ref{table.rvs1296} and \ref{table.rvs1298}.\\

The raw RV time series have a standard deviation of 28.4 and 28.9 m/s for TOI-1296 and TOI-1298, respectively, with a periodic signal. When folded at the period of the photometric transit, the sine wave signal is clearly detected, corresponding to semi-amplitudes of 35.0$\pm$3.5 and 34.4$\pm$2.5 m/s, respectively. After this signal is removed, the periodogram of the RV residuals does not show another signal for any of the systems.

\begin{figure}
\centering
\includegraphics[width=\hsize]{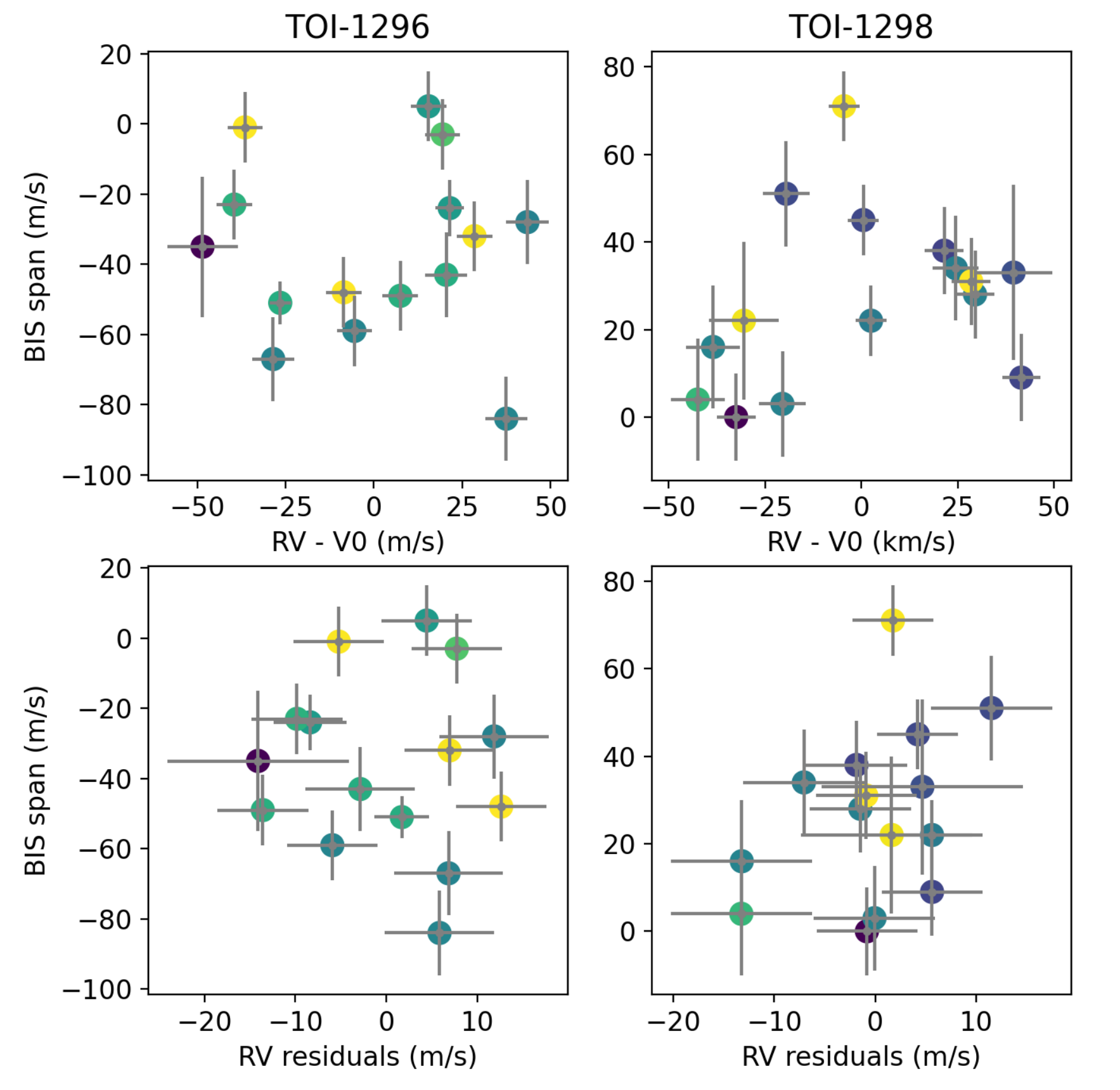}
\caption{
Bisector span as a function of the RV for both stars, and as a function of the residuals after the planet models were modelled. The colours indicate the phase in a rotation cycle. }
\label{BIS}
\end{figure}

On the other hand, there is some variability observed in the bisector span time series at a level of 1.5 to 3$\sigma$, as shown in Figure \ref{BIS}: this figure both shows how the bisector of the line varies as a function of the RV for both stars (top panels) and as a function of the RV residuals when the best-fit planet model is removed. The colour scale indicates the rotation cycle, using the estimated rotation periods given in section 3.1. If the bisector span varies with time, which may be the indication of stellar activity, the correlations are weak. Moreover, this possible activity signal does not occur at the period corresponding to the planet orbital period. Pearson's coefficients are lower than 0.4 in all cases. The greatest correlation value of 0.39 is seen between the RV residuals and bisector span for TOI-1298 (bottom right panel). Stellar activity is thus excluded as the main source of the RV variations, while there may be some activity contribution to the RV residuals for TOI-1298. The RV signal in phase with the TESS ephemeris for both systems establishes the planetary nature of the transiting bodies.

\section{Results}

\begin{figure}
\centering
\includegraphics[width=0.9\hsize]{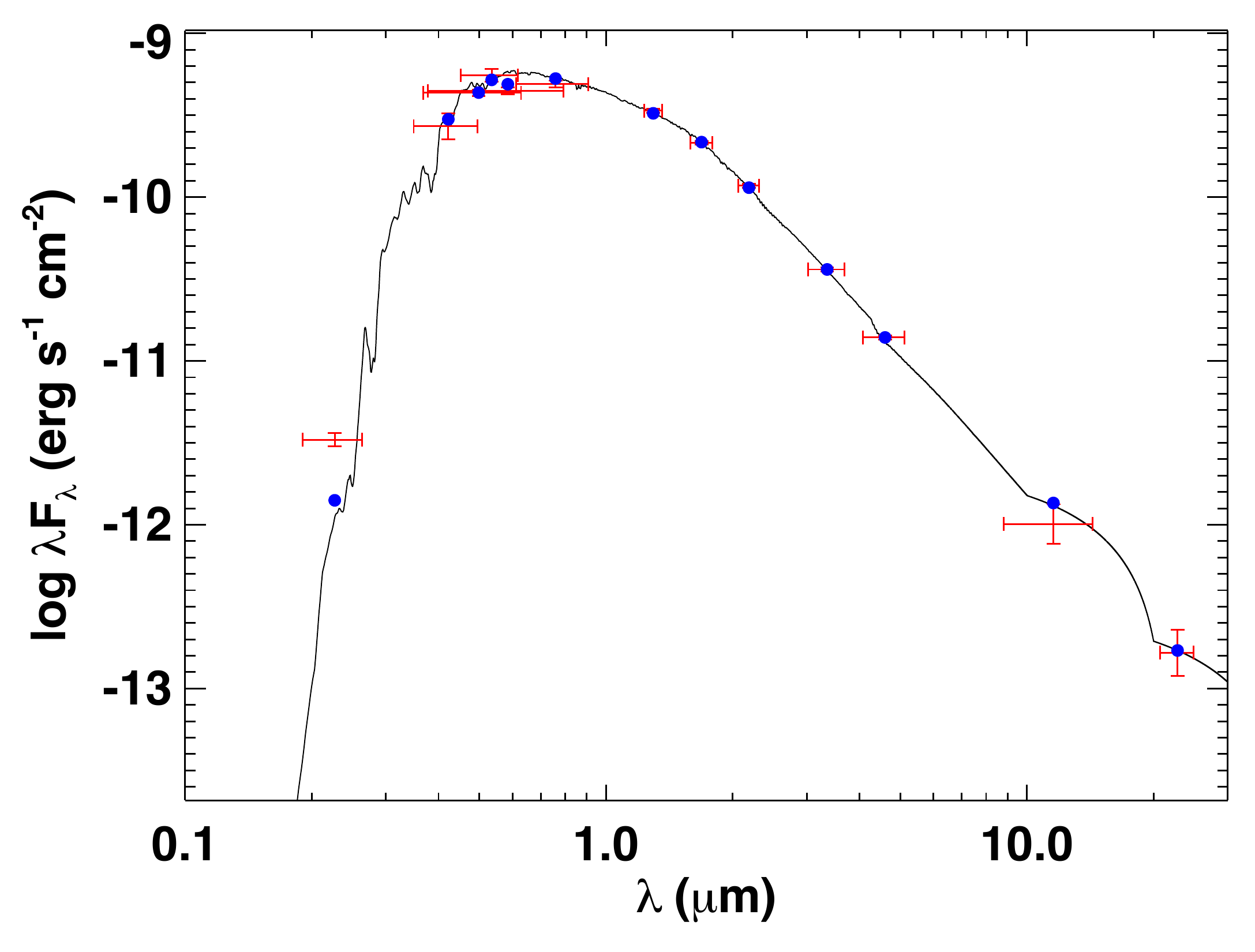}
\includegraphics[width=0.9\hsize]{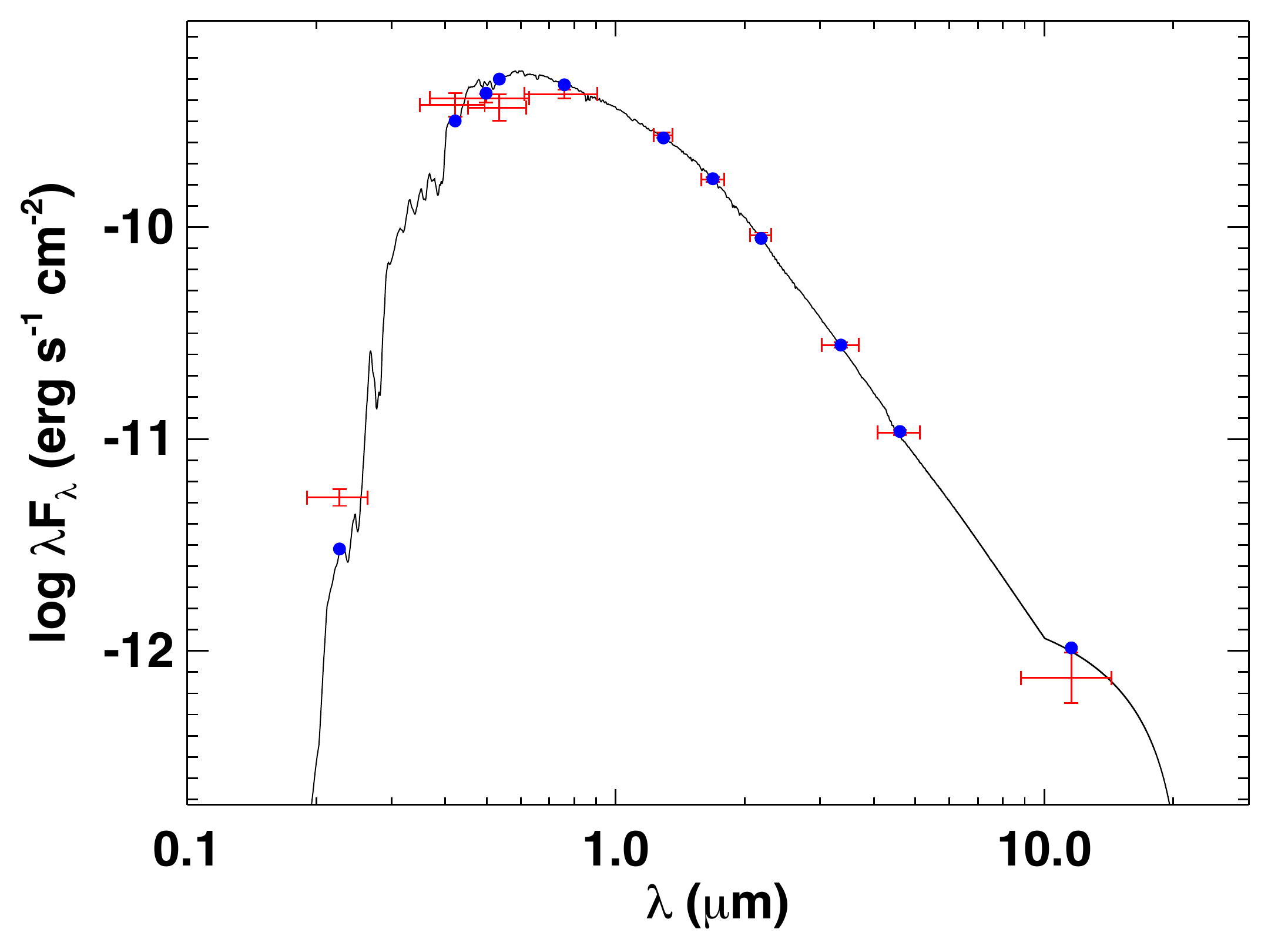}
\caption{Spectral energy distributions of TOI-1296 (top) and TOI-1298 (bottom). Red symbols represent the observed photometric measurements, where the horizontal bars represent the effective width of the passband. Blue symbols are the model fluxes from the best-fit Kurucz atmosphere model (black). 
}
\label{fig:sed}
\end{figure}

\begin{figure}
\centering
\includegraphics[width=0.9\hsize]{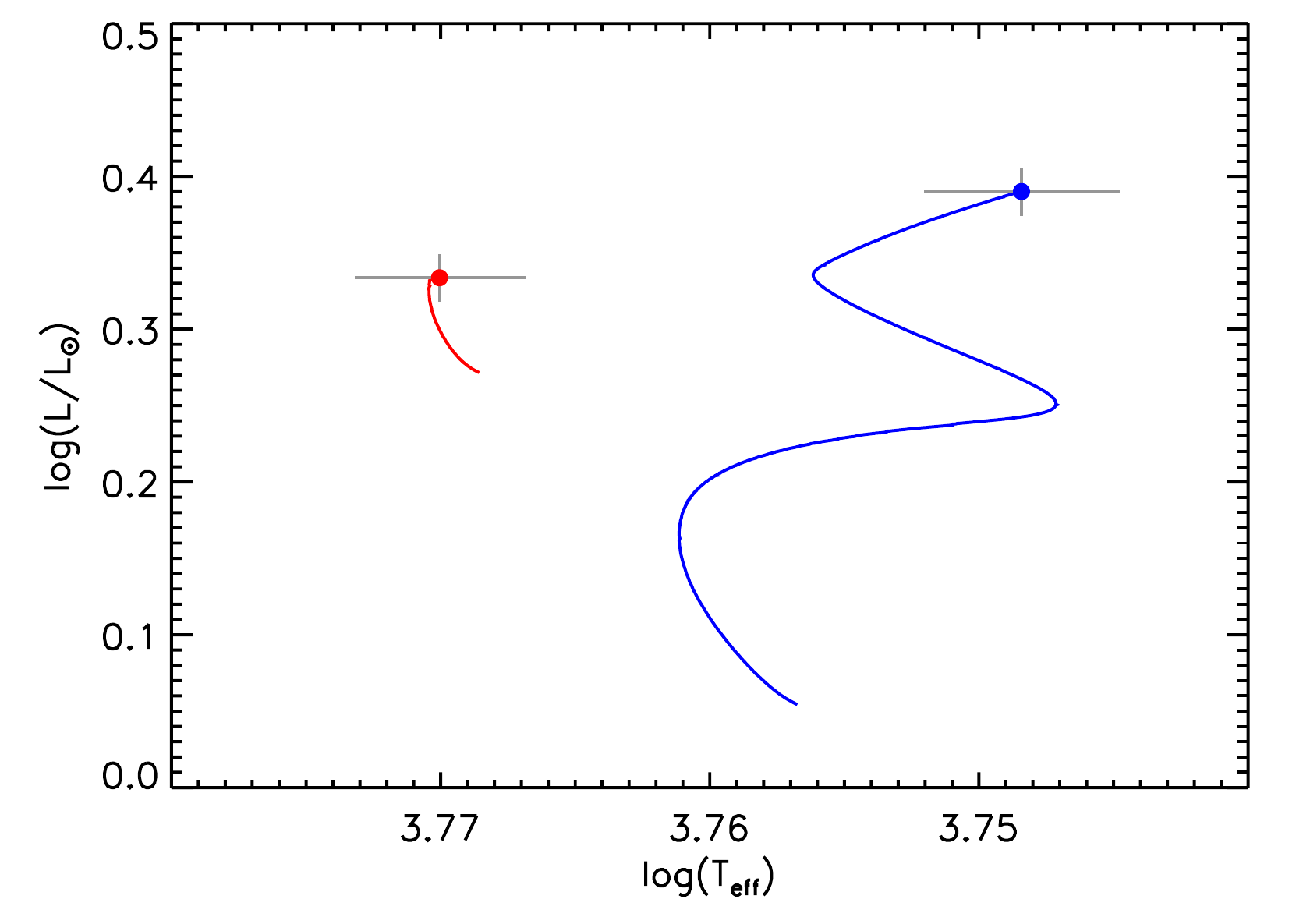}
\caption{ HR diagram in luminosity as a function of effective temperature. The isochrones are from the CESAM2k  models \citep{cesam}.
Symbols show the measured position of TOI-1296 (blue) and TOI-1298 (red) in this diagram, the first one lying on the subgiant branch, and the second one on the main sequence. 
}
\label{fig:hr}
\end{figure}

\subsection{Stellar parameters}\label{sec:stellarparams}
For each star, all spectra not contaminated by the Moon and of sufficient quality were summed up in order to get a stellar template of a high S/N (total S/N of 90 and 110 for TOI-1296 and TOI-1298, respectively). We then used the method described in \citet{santos2013} and \citet{sousa2018}: from the measured equivalent widths of reference lines with ARES, stellar atmospheric parameters and chemical abundances were derived in the local thermodynamic equilibrium with the latest version of the MOOG code \citep{sneden1973} and a grid of Atlas 9 model atmospheres \citep{kurucz1993}. The effective temperature, log$g$, projected rotational velocity $v$sin$i$, turbulence velocity $v_{tur}$, and metallicity values [Fe/H] derived from this spectroscopic analysis are listed in Table \ref{tab:stars}.
The derived spectroscopic parameters are in close agreement with those obtained from the TRES measurements. In particular, the large metallicity value is confirmed with both independent measurements.

We performed an analysis of the broadband spectral energy distribution (SED) of the star together with the {\it Gaia\/} EDR3 parallaxes \citep[with no systematic correction; see, e.g.][]{StassunTorres:2021} in order to determine an empirical measurement of the stellar radius, following the procedures described in \citet{Stassun:2016} and \citet{Stassun:2017,Stassun:2018}. We obtained the $B_T V_T$ magnitudes from {\it Tycho-2}, the $JHK_S$ magnitudes from {\it 2MASS}, the W1--W4 magnitudes from {\it WISE}, the $G G_{\rm BP} G_{\rm RP}$ magnitudes from {\it Gaia}, and the near-ultraviolet (NUV) magnitudes from {\it GALEX}. Together, the available photometry spans the full stellar SED over the wavelength range 0.2--22~$\mu$m (see Figure~\ref{fig:sed}).

We then performed a fit using Kurucz stellar atmosphere models, with $T_{\rm eff}$, [Fe/H], and $\log g$ adopted from the spectroscopic analysis (Table \ref{tab:stars}). The only additional free parameter is the extinction ($A_V$), which we restricted to the maximum line-of-sight value from the dust maps of \citet{Schlegel:1998}. The resulting fits for TOI-1296 and TOI-1298 are good (Figure~\ref{fig:sed}) with a reduced $\chi^2$ of 1.6 and 2.0, respectively, excluding the NUV flux which appears slightly in excess and could indicate modest chromospheric activity. The best-fit extinction values are $A_V = 0.05 \pm 0.05$ for TOI-1296 and $0.04 \pm 0.04$ for TOI-1298. Integrating the (unreddened) model SED gives the bolometric flux at Earth, $F_{\rm bol} = 7.81 \pm 0.27 \times 10^{-10}$ erg~s$^{-1}$~cm$^{-2}$ and $6.89 \pm 0.24 \times 10^{-10}$ erg~s$^{-1}$~cm$^{-2}$ for TOI-1296 and TOI-1298, respectively, which with the parallax directly gives the bolometric luminosity $L_{\rm bol} = 2.455 \pm 0.088$~L$_\odot$ and $2.156 \pm 0.077$~L$_\odot$, respectively. Taking the $F_{\rm bol}$ and $T_{\rm eff}$, together with the {\it Gaia\/} parallax, gives the stellar radius $R_\star = 1.664 \pm 0.041$~R$_\odot$ and $1.412 \pm 0.033$~R$_\odot$, respectively.

We can also infer the ages of the systems using empirical activity-age relationships. In particular, the observed UV excess in Figure~\ref{fig:sed} via the empirical relations of \citet{Findeisen:2011} yields an activity index $\log R'_{\rm HK} = -5.49 \pm 0.1$ and $-5.57 \pm 0.1$, which implies a rotation period of $P_{\rm rot} = 54 \pm 5$~d and $38 \pm 3$~d for TOI-1296 and TOI-1298, respectively. There is no significant sign of these rotation periods in the TESS light curves. Adopting this activity with the empirical rotation-age relations of \citet{Mamajek2008} gives age estimates of 10~Gyr (TOI-1296) and 9.5~Gyr (TOI-1298).

To better determine the stellar parameters of the two targets, we performed an optimisation using the stellar evolution code \textsc{CESAM2K} (\citealt{cesam}) within a Levenberg-Marquardt algorithm. We used the estimates of $T_{\rm eff}$, $\log g$, $[$Fe/H$]$, and $L_{\rm bol}$ derived in this study as observational constraints, and we considered the stellar mass, the age, the initial metallicity, and the initial helium abundance as free parameters of the fit. In the models, we used the OPAL 2005 equation of state (\citealt{rogers02}) and opacity tables. The nuclear reactions rates were computed using the NACRE compilation \citep{angulo99} and the LUNA revised rate for the $^{14}$N$(p,\gamma)^{15}$O reaction  \citep{formicola04}. Convection was treated following the formalism of \cite{canuto96} with a solar-calibrated value of the mixing length parameter. Core overshooting was neglected. Microscopic diffusion was included by solving the equations of \cite{burgers69}, but the effects of radiative levitation were neglected. The atmosphere was described by Eddington’s grey law. We adopted the solar mixture of heavy elements of \cite{asplund09}.

The optimal stellar parameters obtained for the two stars are given in Table \ref{tab:stars}, and the evolutionary tracks of the best models in the HR diagram are shown in Fig. \ref{fig:hr}. We found TOI-1296 to be in the subgiant phase, while TOI-1298 still lies on the main sequence, the optimal model having a fraction of hydrogen in the core of $X_{\rm c} = 0.51$. The error bars given in Table \ref{tab:stars} were obtained by using the inverse of the Hessian matrix. They correspond to internal error bars and do not account for the effects of systematics caused by our choice of input physics. 

\subsection{Data modelling}

We performed a global modelling including all available data. The fitted parameters are as follows: $q_1$ and $q_2$ \citep{kipping2013}, the quadratic limb darkening coefficients for all light curves, the radius to semi-major axis ratio, the radius ratio, the orbit inclination, the mid-transit time, the orbital period, the semi-amplitude of RV variations, the systemic velocity, the stellar density, and a jitter term for the RV time series.
Eccentricity was let free to vary, adding the fitted parameters $\sqrt{e}\cos\omega$ and $\sqrt{e}\sin\omega$.
For TOI-1296b, Gaussian process regression was used for TESS data and 
all ground-based light curves, which were modelled simultaneously with SOPHIE data.
For TOI-1298b, only TESS and SOPHIE data were used in the modelling.
The TESS data included in the analysis are centred on the transit events, with a width of three times their duration. 

The priors used in the analysis are uniform for most of the parameters, except the stellar density for which a normal prior is used from the stellar mass and radius derived from the spectroscopic analysis and SED modelling, as discussed in section~\ref{sec:stellarparams}. The priors we imposed on the stellar density are the following: $0.358\pm0.052~\rm{g\;cm^{-3}}$ for TOI-1296 and
$0.721\pm0.074~\rm{g\;cm^{-3}}$ for TOI-1298.

For this global modelling, we used the \texttt{Juliet} package \citep{espinoza2019}. Within \texttt{Juliet}, we used the transit model \texttt{batman} \citep{kreidberg2015} and the
RV model \texttt{radvel} \citep{fulton2018}.
We used the approximate Matern kernel included in \texttt{celerite} \citep{foreman-mackey2017} for the photometric datasets, including both TESS data and ground-based light curves.
Quadratic limb darkening was let free for each filter.
We oversampled the TESS 30-minute cadence with a factor of 30, and the TESS 2-minute cadence with a factor of 3 \citep{kipping2010}.
To sample from the posterior, we used the nested sampling code \texttt{dynesty} \citep{speagle2020}, with 500 live-points and a relative value of difference of the logarithm of the evidence less than $0.5$ as convergence criterium.

\begin{table*}
\renewcommand{\arraystretch}{1.25}
\centering
\caption{Posterior median and 68.3\% credible interval for the system's parameters from the \texttt{Juliet} analysis.}\label{table.juliet}
\begin{tabular}{lccc}
\hline
Parameter & Units & TOI-1296 & TOI-1298\\
\hline
Stellar mean density                          & [$\rm{g\;cm^{-3}}$] & 0.324$_{-0.048}^{+0.037}$       & 0.670 $\pm$ 0.062 \\
$q_1$ TESS                                    &                     & 0.354$_{-0.075}^{+0.092}$       & 0.180$_{-0.076}^{+0.13}$  \\
$q_2$ TESS                                    &                     & 0.300$_{-0.077}^{+0.087}$       & 0.44$_{-0.21}^{+0.29}$  \smallskip\\
Orbital period, P                             & [days]              & $3.9443715 \pm 5.8\;10^{-6}$    & $4.537164 \pm 1.2\;10^{-5}$ \\
Mid-transit time, Tc                          & [BJD$_{\rm TDB}$]   & 2458930.75532 $\pm$ 0.00019     & 2458929.58558 $\pm$ 0.00031 \\
Semi-major axis in stellar radii, $a/R_\star$ &                     & 6.44$_{-0.33}^{+0.24}$          & 9.00$_{-0.29}^{+0.25}$  \\
Semi-major axis, $a$                          & [au]                & 0.0497$_{-0.0028}^{+0.0023}$    & 0.0590 $\pm$ 0.0023  \\
Radius ratio, $R_{\mathrm{p}}/R_\star$        &                     & 0.07599$_{-0.00039}^{+0.00046}$ & 0.06119 $\pm$ 0.00053 \\
Impact parameter, $b$                         &                     & 0.13 $\pm$ 0.11                 & 0.16$_{-0.11}^{+0.13}$ \\    
Orbital inclination, $i$  [degrees]      &       & 88.81$_{-1.0}^{+0.82}$    & 88.96$_{-0.86}^{+0.73}$ \\
$\sqrt{e}\cos{\omega}$                        &                     & -0.12$_{-0.14}^{+0.18}$         & -0.05$_{-0.12}^{+0.14}$ \\
$\sqrt{e}\sin{\omega}$                        &                     & 0.11$_{-0.17}^{+0.15}$          & 0.04$_{-0.15}^{+0.13}$ \\ 
Eccentricity, $e$                             &                     & 0.055$_{-0.038}^{+0.061}$       & 0.032$_{-0.023}^{+0.034}$ \\    
Argument of pericentre, $\omega$              & [degrees]           & 137 $\pm$ 62                    & 130 $\pm$ 92 \\     
Radial velocity semi-amplitude, $K$           & [$\rm{m\;s^{-1}}$]  & 34.8 $\pm$ 3.4                  & 34.4 $\pm$ 2.5 \\
Systemic velocity, $\mu$                      & [$\rm{km\;s^{-1}}$] & 25.3919 $\pm$ 0.0026            & -55.5587 $\pm$ 0.0025 \\
Jitter radial velocity, $\sigma_{\rm RV}$     & [$\rm{m\;s^{-1}}$]  & 7.7$_{-2.3}^{+3.0}$             & 4.0$_{-3.8}^{+3.4}$ \smallskip\\
Planet mass, $M_{\rm p }$                     & [\MJ]               & 0.298 $\pm$ 0.039               & 0.356 $\pm$ 0.032 \\
Planet radius, $R_{\rm p}$                    & [\RJ]               & 1.231 $\pm$ 0.031               & 0.841 $\pm$ 0.021 \\
Planet mean density, $\rho_{\rm p}$           & [$\rm{g\;cm^{-3}}$] & 0.198 $\pm$ 0.031               & 0.743 $\pm$ 0.091 \\
Planet equilibrium temperature$^\dagger$, T$_{\rm eq}$ & [K]        & 1562$_{-31}^{+43}$              & 1388 $\pm$ 24 \smallskip\\
\hline
\end{tabular}
\begin{list}{}{}
\item {\bf{Notes.}}
$^\dagger$ For zero albedo and full day-night heat redistribution.
\end{list}
\end{table*}

For the system TOI-1296, the posterior distribution value for the stellar density we obtained, including all available data, is $0.324_{-0.048}^{+0.037}$~g/cm$^3$. With this good agreement on stellar density values, we safely adopted the radius and mass from the stellar characterisation using the SED modelling. The planet's mass and radius we derived are as follows: $M_{pl} = 0.298\pm0.039$~\MJ\ and $R_{pl} = 1.231\pm0.031$~\RJ. This corresponds to a planet's density $\rho_p$ of $0.198\pm0.031$~g/cm$^3$. The derived eccentricity is compatible with a circular orbit.\\

\begin{figure*}
\centering
\includegraphics[width=1.0\hsize]{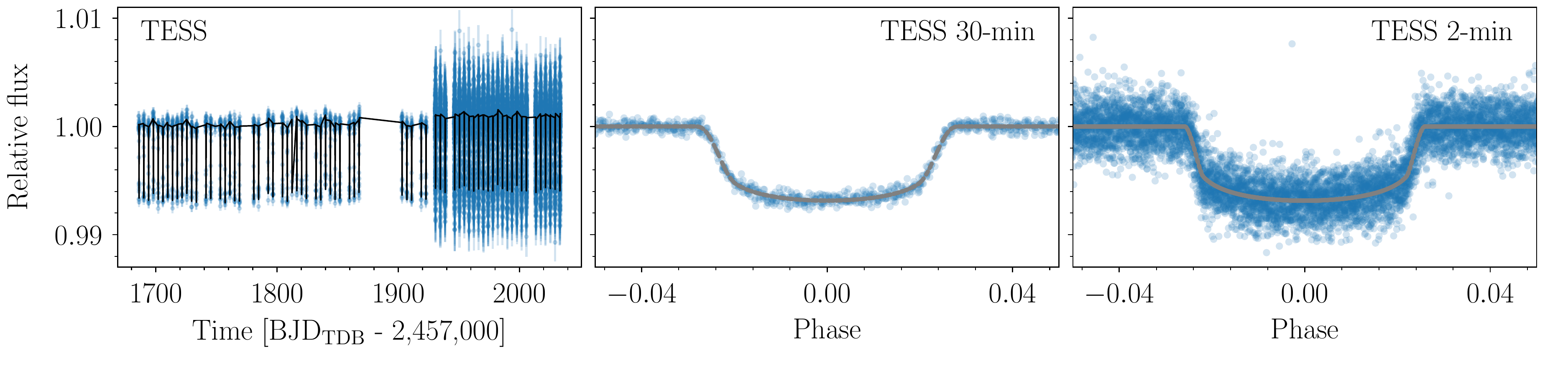}
\includegraphics[width=1.0\hsize]{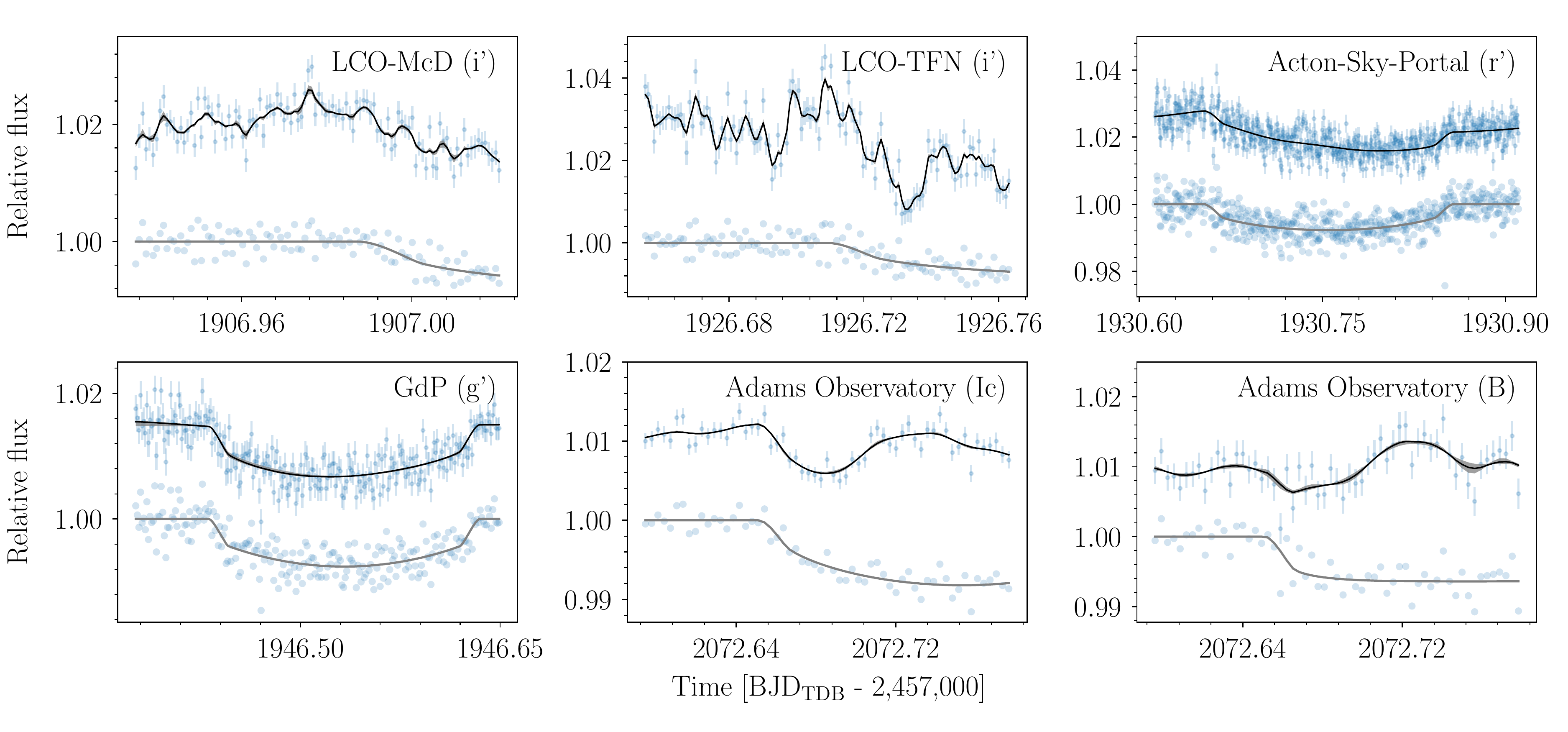}
\includegraphics[width=0.49\hsize]{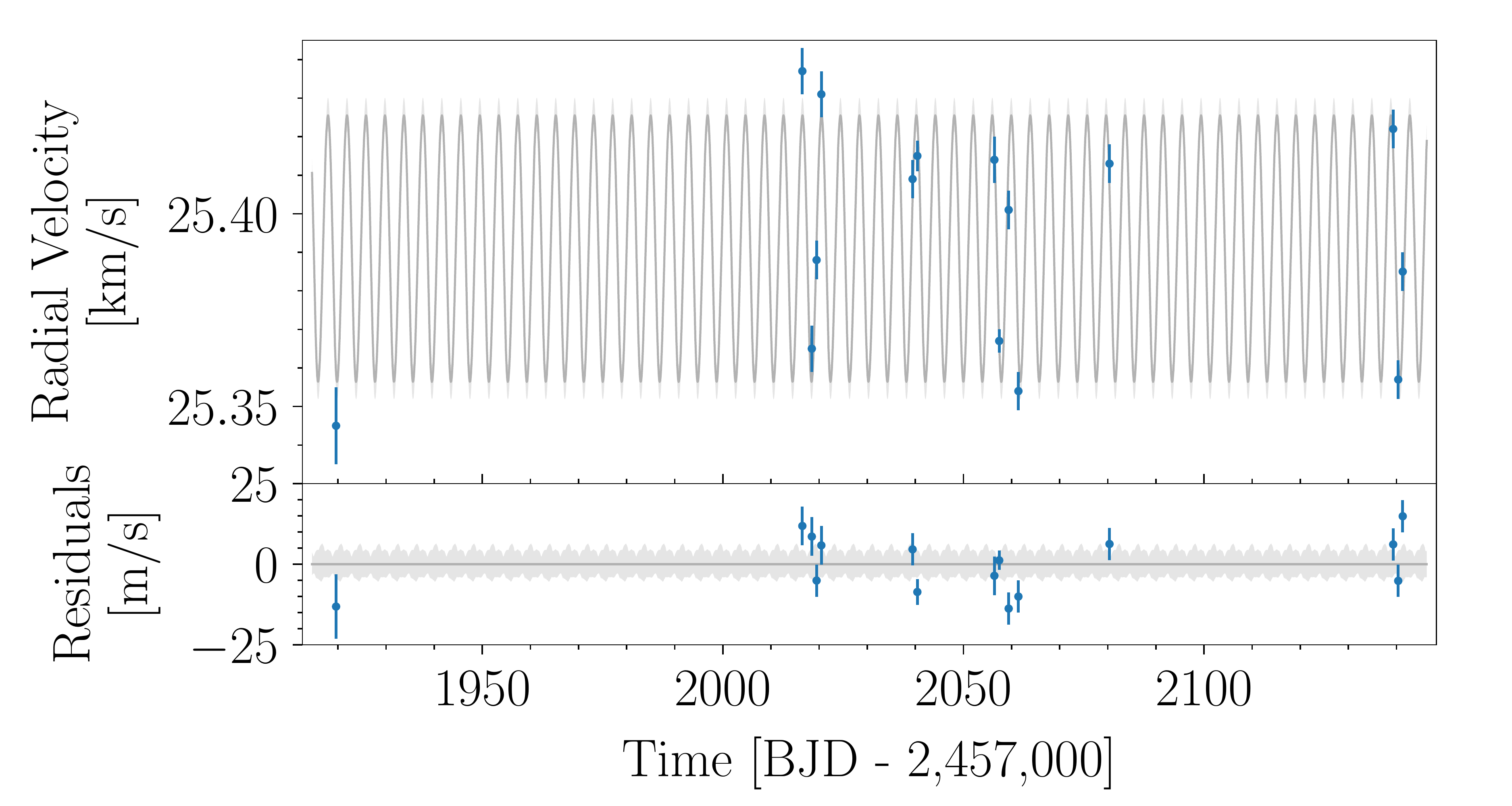}
\includegraphics[width=0.49\hsize]{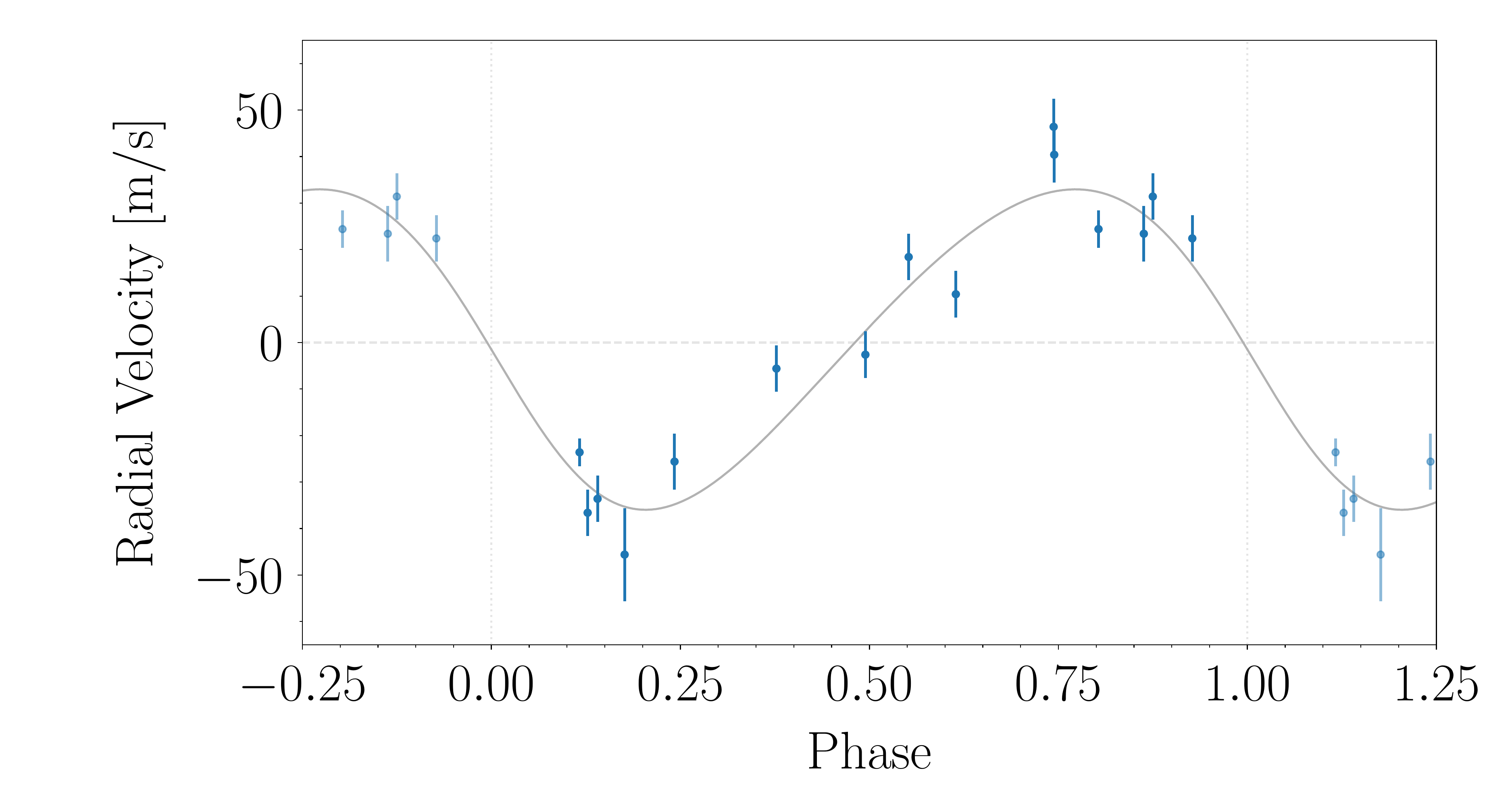}
\caption{ Modelling of data obtained on TOI-1296b. (top) The TESS light curve in 30-minute and 2-minute cadence, and ExoFOP ground-based light curves; (bottom) The SOPHIE data series of RV versus time and folded RV data, together with the best-fit model (the time unit for the RV time series is BJD$_{UTC}$}). In all panels, blue data points with error bars show the observed data. Black lines and intervals in grey show the model median and the 68.3\% credible interval computed from 1000 random posterior samples. Grey lines show the maximum a posteriori transit model. The blue points associated with those grey lines are the data points corrected for the best-fit Matern kernel. 
For the ground-based light curves, the observatory name and filters used for observations label the relevant plot. 

\label{TOI1296-fit}
\end{figure*}

\begin{figure*}
\centering
\includegraphics[width=1.0\hsize]{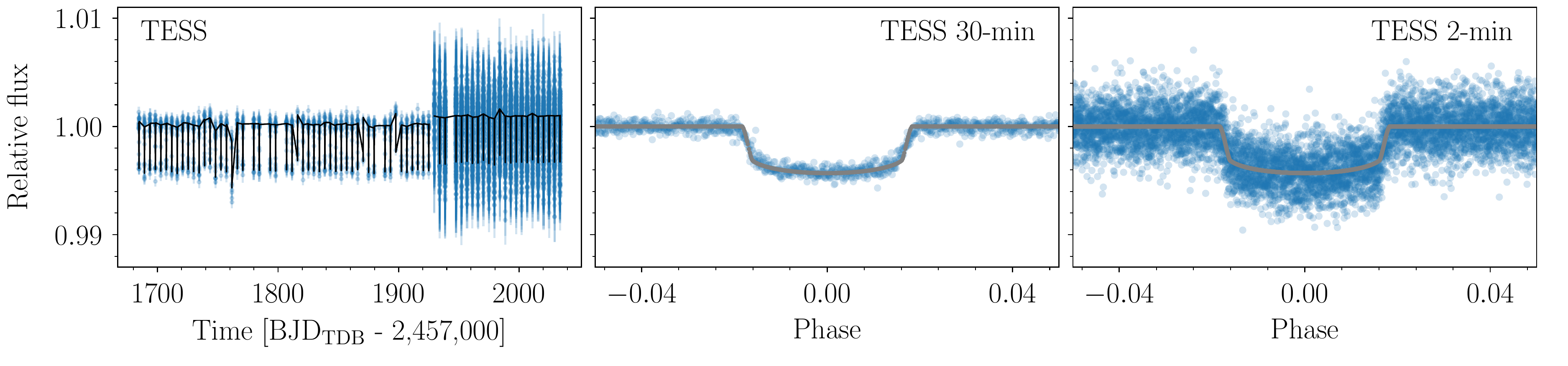}
\includegraphics[width=0.49\hsize]{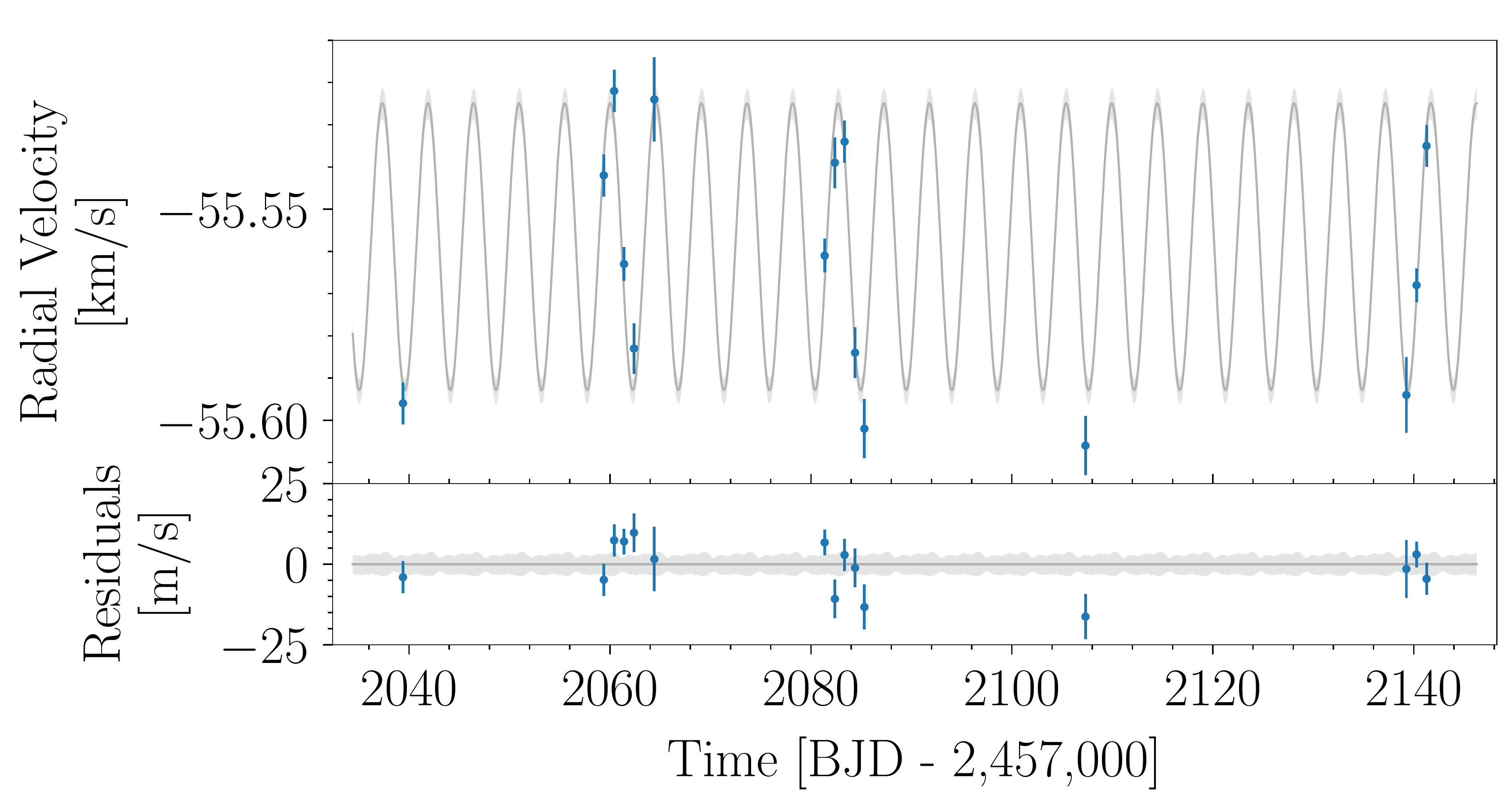}
\includegraphics[width=0.49\hsize]{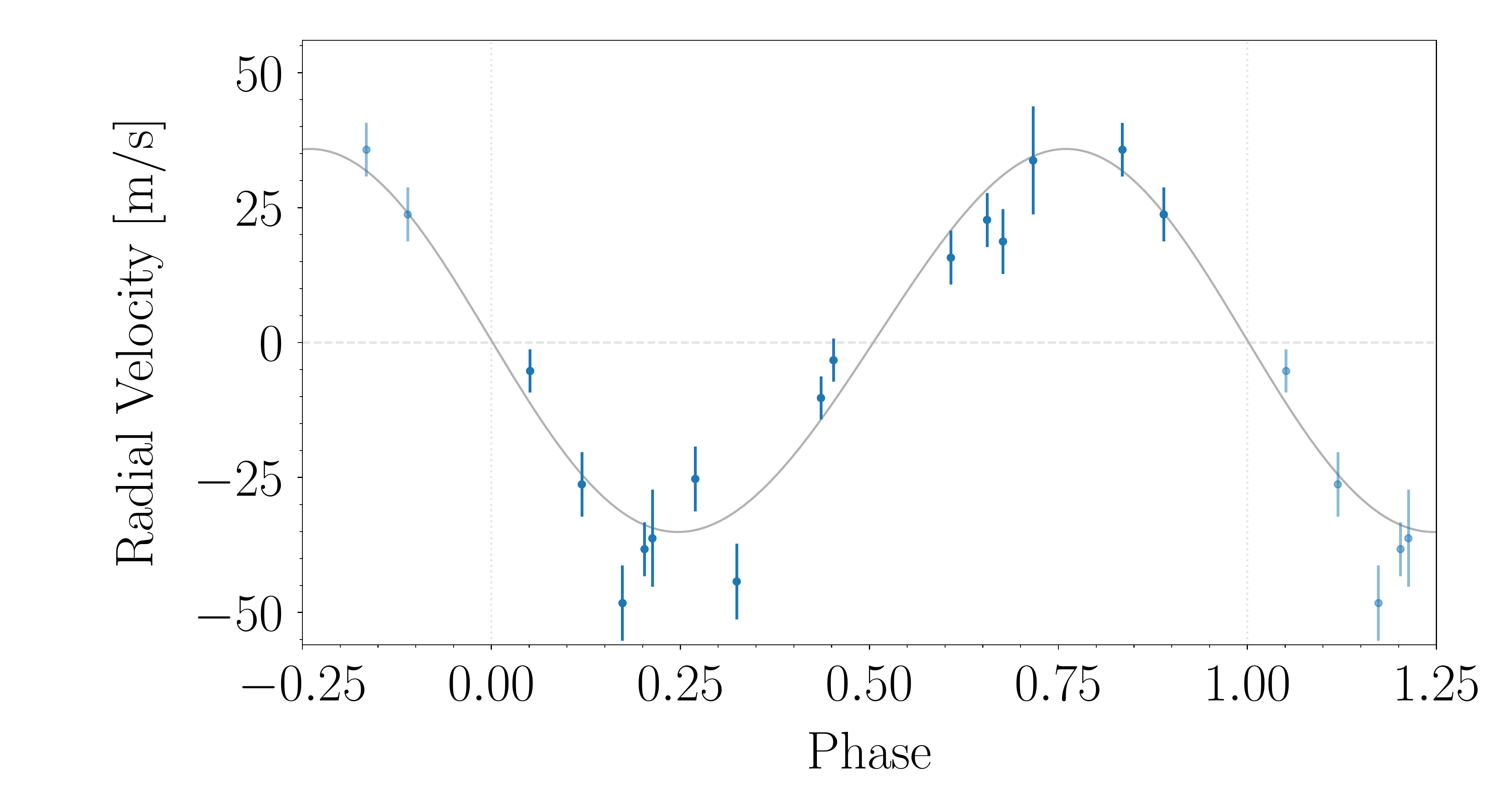}
\caption{Same as Figure \ref{TOI1296-fit}, but for TOI-1298b.}
\label{TOI1298-fit}
\end{figure*}

TOI-1298 was modelled in the same way. 
We obtain a posterior value of $0.670\pm0.062~\rm{g\;cm^{-3}}$, showing good agreement with the prior value of $0.721\pm0.074\rm{g\;cm^{-3}}$. The derived planet's mass is $0.356\pm0.032$~\MJ\ and the radius is $0.841\pm0.021$~\RJ, implying a density of $0.743\pm0.091$~g/cm$^3$. The eccentricity is not significantly different from 0. 
The posterior results are listed in Table~\ref{table.juliet} for TOI-1296 and TOI-1298 systems, while Figures \ref{TOI1296-fit}  and \ref{TOI1298-fit} show the best fit to the data in both cases. 

\begin{figure}
\centering
\includegraphics[width=0.95\hsize]{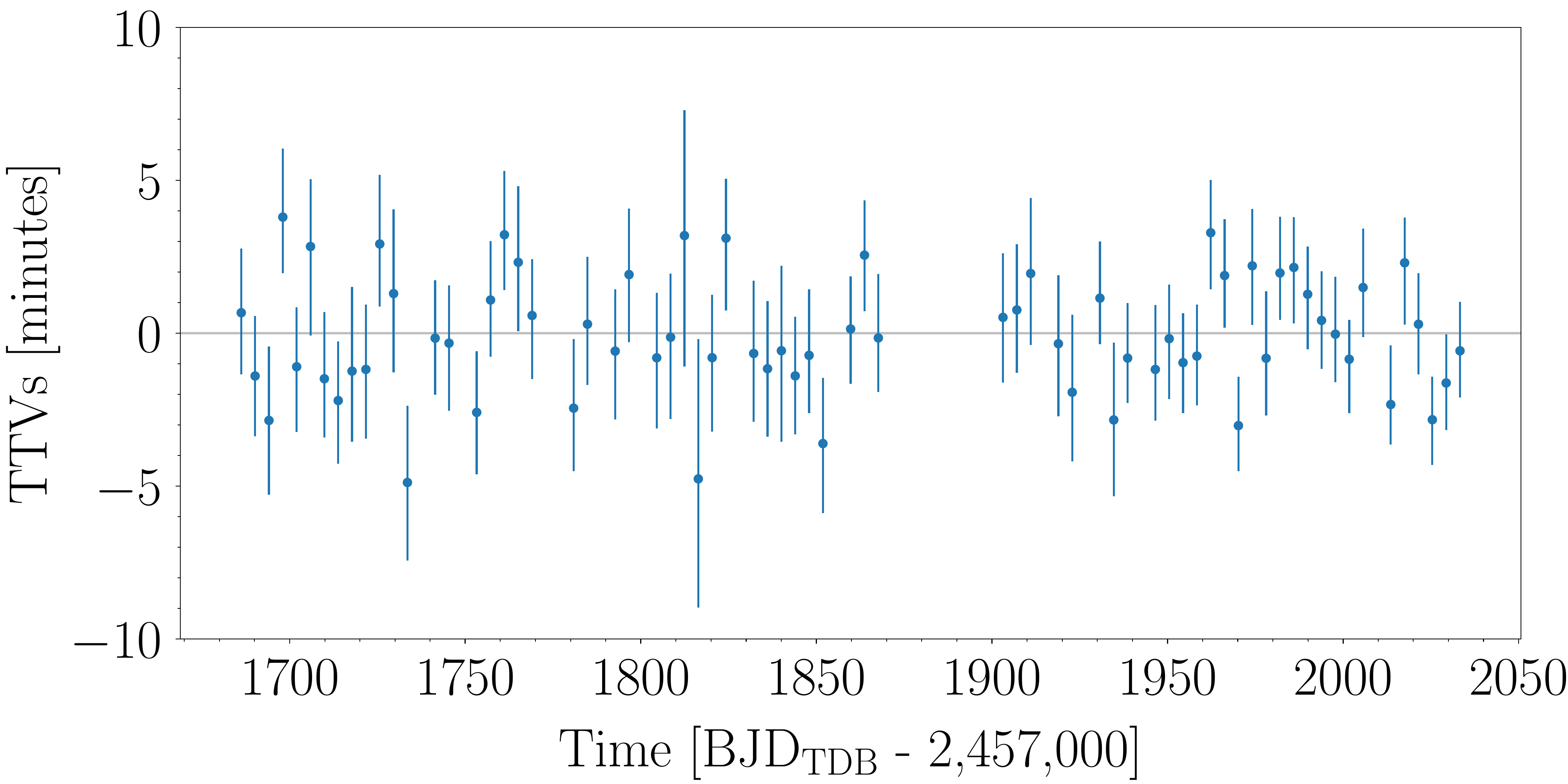}
\includegraphics[width=0.95\hsize]{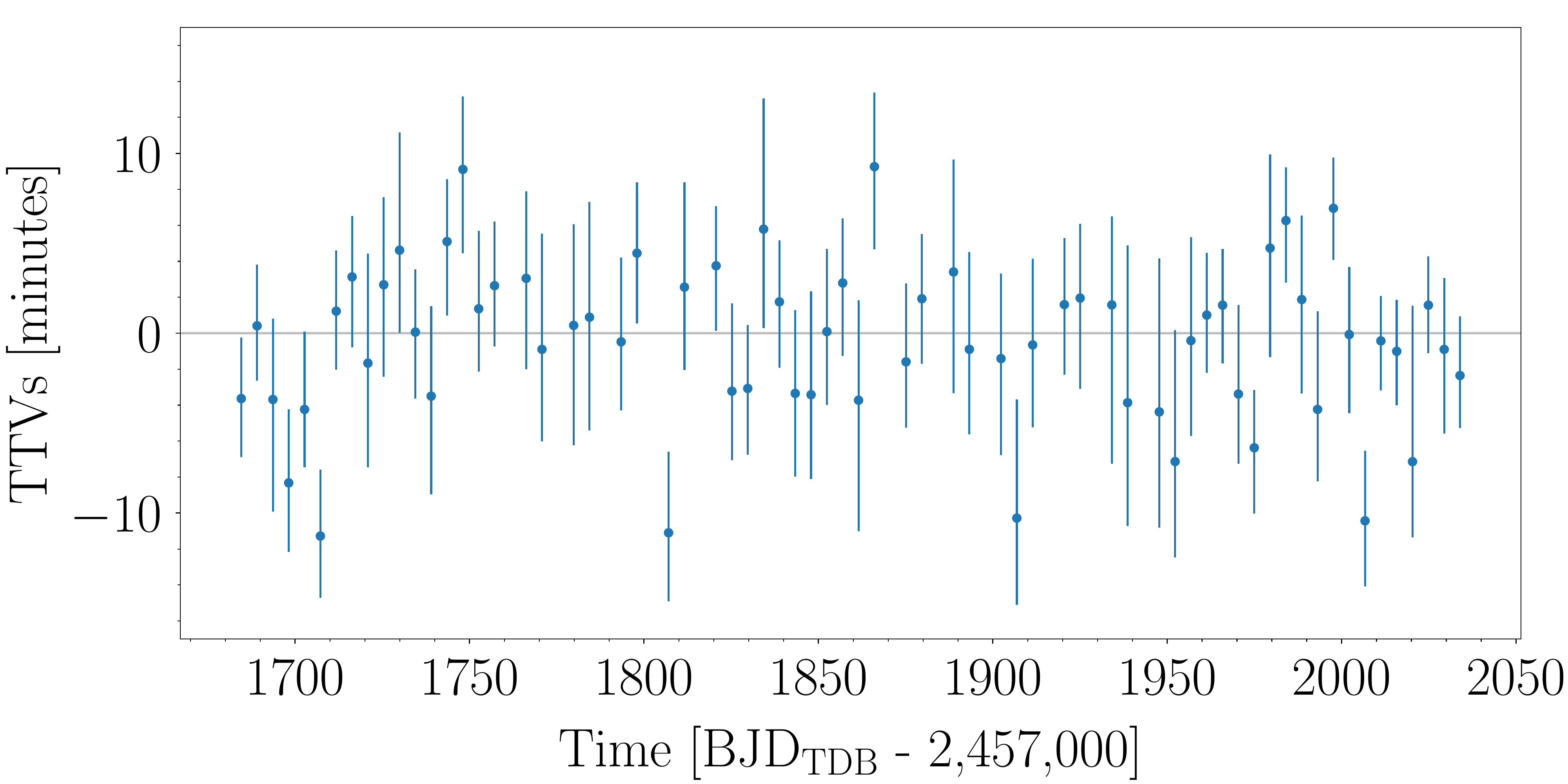}
\caption{Deviation of individual transit times with respect to linear ephemeris, as a function of time (top) for TOI-1296b and (bottom) for TOI-1298b.}
\label{TTV}
\end{figure}

We then fixed the transit parameters to the posterior maximum and let the time of individual transits vary to search for transit time variations. We used a normal prior with a width of 0.03 days with respect to the linear ephemeris of the fit, obtained in the global modelling. The resulting transit times are shown in Figure \ref{TTV} for both systems. It is intriguing that four negative outliers stand out for TOI-1298b at regular times, separated by 100 days. It may be due to an instrumental systematic noise, as no global pattern is seen at this period. We conclude that there is no significant detection of variable transit times for either of the systems.

\section{Discussion}
The planets discovered by TESS around TOI-1296 and TOI-1298 both have short orbital periods of 3.94 and 4.54~days, respectively. They are both heavily irradiated by their host star with estimated equilibrium temperatures of about 1560 and 1390~K and incident fluxes of $\sim$ 7$\times$10$^8$ and 3$\times$10$^8$ erg/s/cm$^2$. The numerous transits revealed by TESS on these systems allowed for a precise radius for the planets, with a large difference between both values of 1.231$\pm0.031$ and 0.841$\pm0.021$ \RJ \ for TOI-1296b and TOI-1298b, respectively. The measured radii of both planets correspond to the bulk of giant-planet radii in the same range of incident flux and thus may be explained by the effects of atmospheric circulation induced by stellar irradiation, as shown by \citet{tremblin2017}. Complementary observations with SOPHIE have established the planetary nature of the signal at this period and secured a mass measurement for both planets in circular orbits. Despite their different radius, both planets have similar masses, close to Saturn's mass. These two planets thus have particularly distinct bulk densities of 0.198 and 0.743~g/cm$^3$. The (slightly) more massive of the two has a much lower radius value. 

It is interesting to compare both stellar host stars. While TOI-1298 still lies on the main sequence, TOI-1296 has started evolving into the subgiant branch, with an enlarged stellar radius (1.66 \RS) and increased luminosity (2.455 \LS). These different evolutionary statuses of both stars may play a role in the radius inflation of the planet TOI-1296b as compared to TOI-1298b, for a mass in a close range, as postulated by \citet{lopez2016}. This type of possible trend between stellar evolution and reinflation of planets was also recently reported by several studies of individual systems such as HAT-P-65, HAT-P-66 \citep{hartman2016}, NGTS-13 \citep{grieves2021}, HD 221416 \citep{huber2019}, and HD 1397 \citep{brahm2019}. The rich harvest of precision transit photometry such as achieved by TESS allows for the discovery and accurate characterisation of large samples of planets that offer observational constraints to evolution models. In the present work, two more planets with 3\% precision on the radius have been presented and their significant difference in terms of their radius value also puts into questions the role of stellar evolution towards the giant branch in the evolution of the close-in planet radius.

\begin{figure}
\includegraphics[width=\hsize]{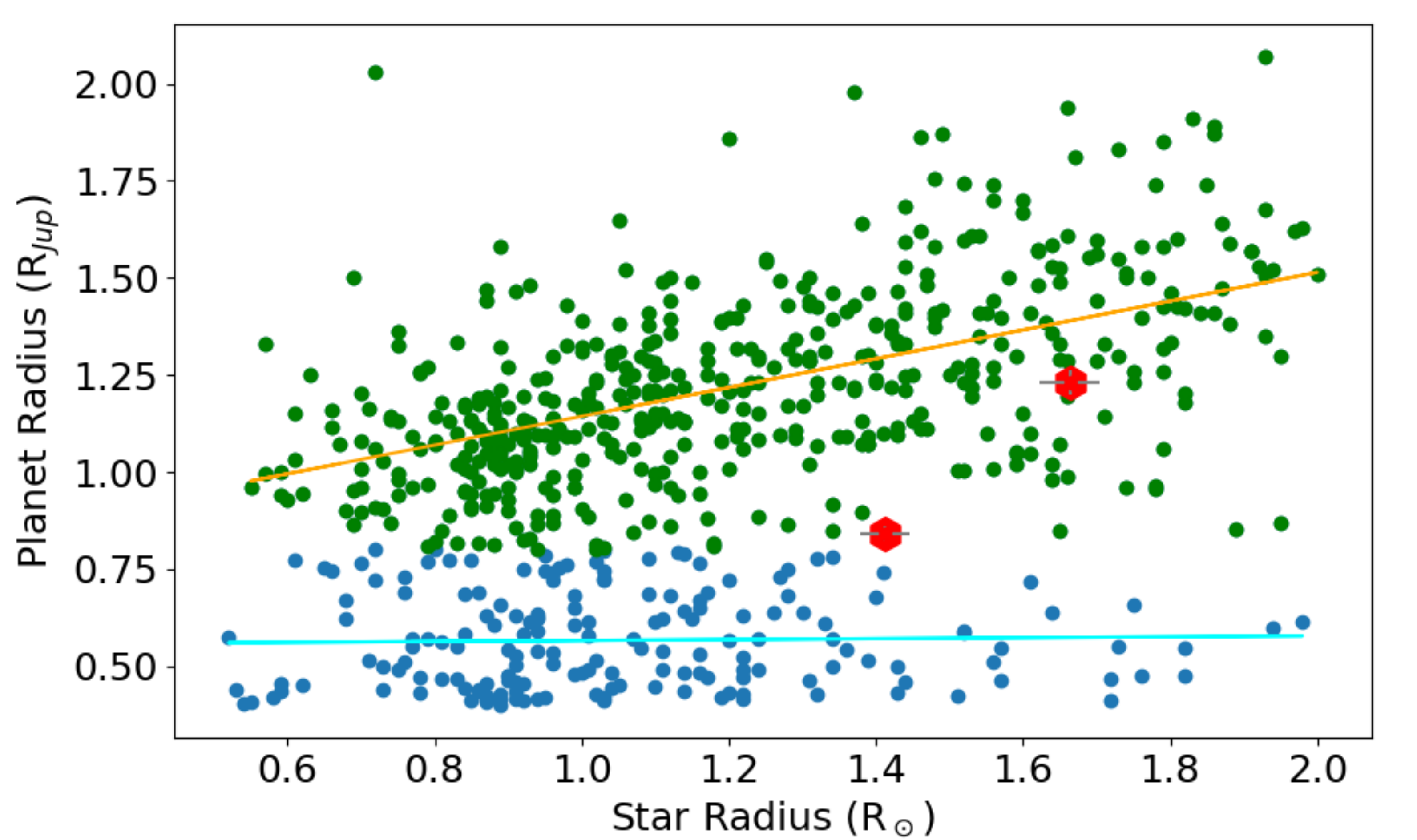}
\includegraphics[width=\hsize]{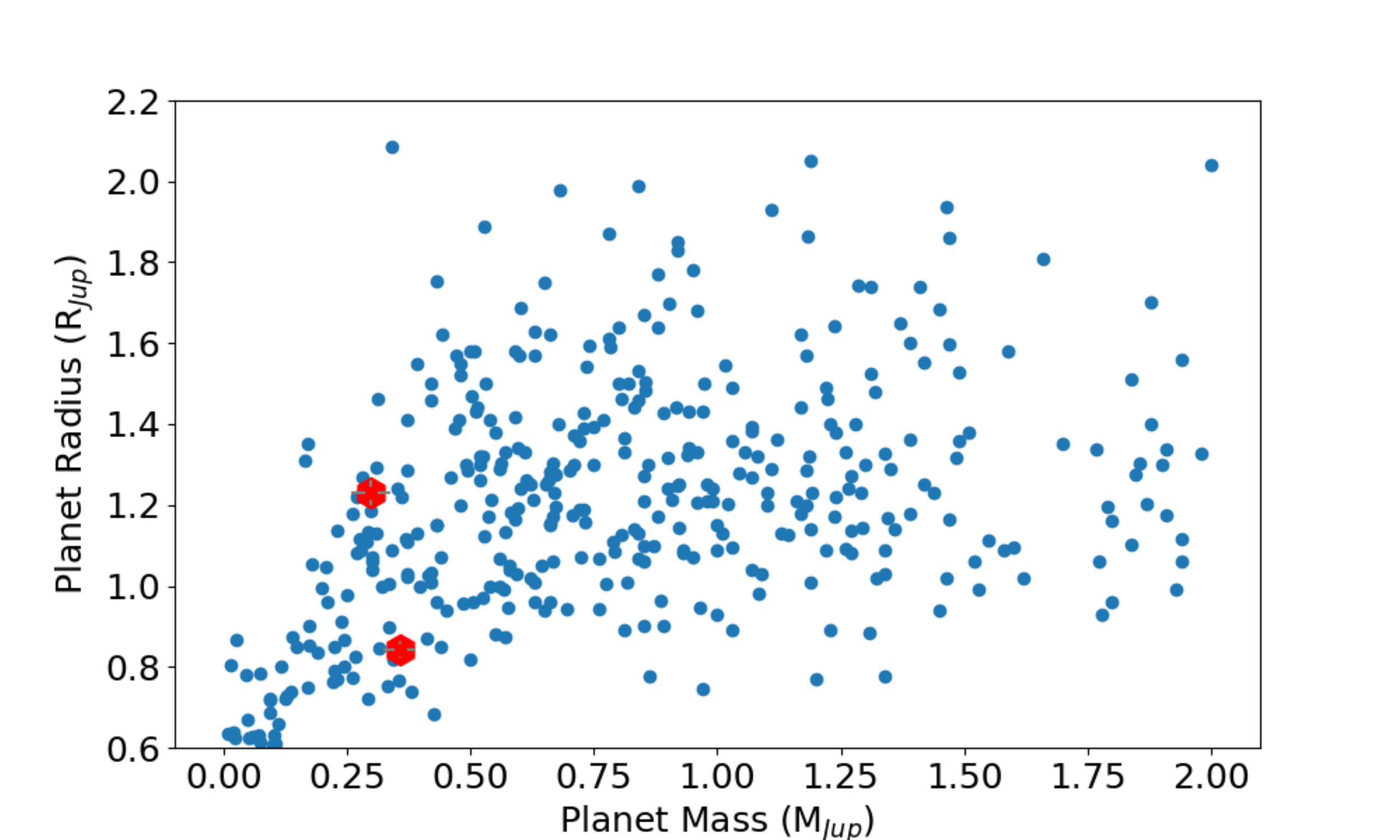}
\caption{ (Top) Planet radius as a function of stellar radius for all transiting giant planets. The two lines are linear fits for the populations of planets above (orange line, green dots) and below (cyan line, blue dots) the 0.8 \RJ\ threshold (see text), respectively. (Bottom) Planet radius as a function of planet mass for transiting giant planets. Small dots represent all confirmed giant planets above 0.5 \RJ\ from the NASA exoplanet archive\protect\footnotemark. The new planets in these parameter spaces are shown with the red symbols. Although of a similar mass, they differ significantly in radius.}
\label{fig:rprs}
\end{figure}

Figure \ref{fig:rprs} (bottom) locates the planets TOI-1296b and TOI-1298b in the mass-radius diagram and shows that, at their given mass, they encompass the whole range of observed planet radii. The top panel of Figure \ref{fig:rprs} shows the current distribution of planetary and stellar radii in the population of confirmed planets. For giant planets with a size between 0.5 and 0.8 \RJ\ there is no visible trend with respect to the stellar radius (cyan line). In contrast, for planets greater than 0.8 \RJ, there is a linear trend between the stellar and planetary radii, with a residual standard deviation of 0.2 \RJ. The radii of TOI-1296 and its giant planet are compatible with this trend within 1$\sigma$. The radius of TOI-1298b, on the other hand, being near the limit where the trend is no longer visible, seems to stand out of the population of small-sized giant planets where the radius does not depend on the stellar radius or lies in a radius valley in between the two populations mentioned above. It is out of the scope of this paper to investigate whether the paucity of planets in the lower right corner of this plot (stellar radii between 1.3 and 2 R$_\odot$ and planetary radius less than 1 \RJ) is an observational bias or directly relates with the evolutionary status of the host stars. No observational bias, however, directly comes into mind, as smaller planets are detected around large-radius stars. It thus appears that when giant planets are present around stars with radii larger than 1.5 R$_\odot$, then their radius increases with the stellar radius. As the process likely relies on received stellar irradiation, it can be related to atmospheric circulation \citep{tremblin2017}. The planet-star distance may explain the dispersion around the trend. The coefficients of this trend are such that R$_p$/\RJ\ = 0.77 + 0.37 $R_s$/$R_\odot$, where R$_p$ and R$_s$ are the planet and stellar radii, respectively. This loose correlation is  similar to the one found between the planet radius and stellar age in \citet{hartman2016} (their Figure 12), and it possibly has the same origin. There is no doubt that current transit surveys and their follow-up will further unveil the properties of giant planets as the host star evolves, especially when the stellar radius can be precisely characterised with asteroseismology, as in \citet{huber2019}.

\begin{figure}
\includegraphics[width=0.9\hsize]{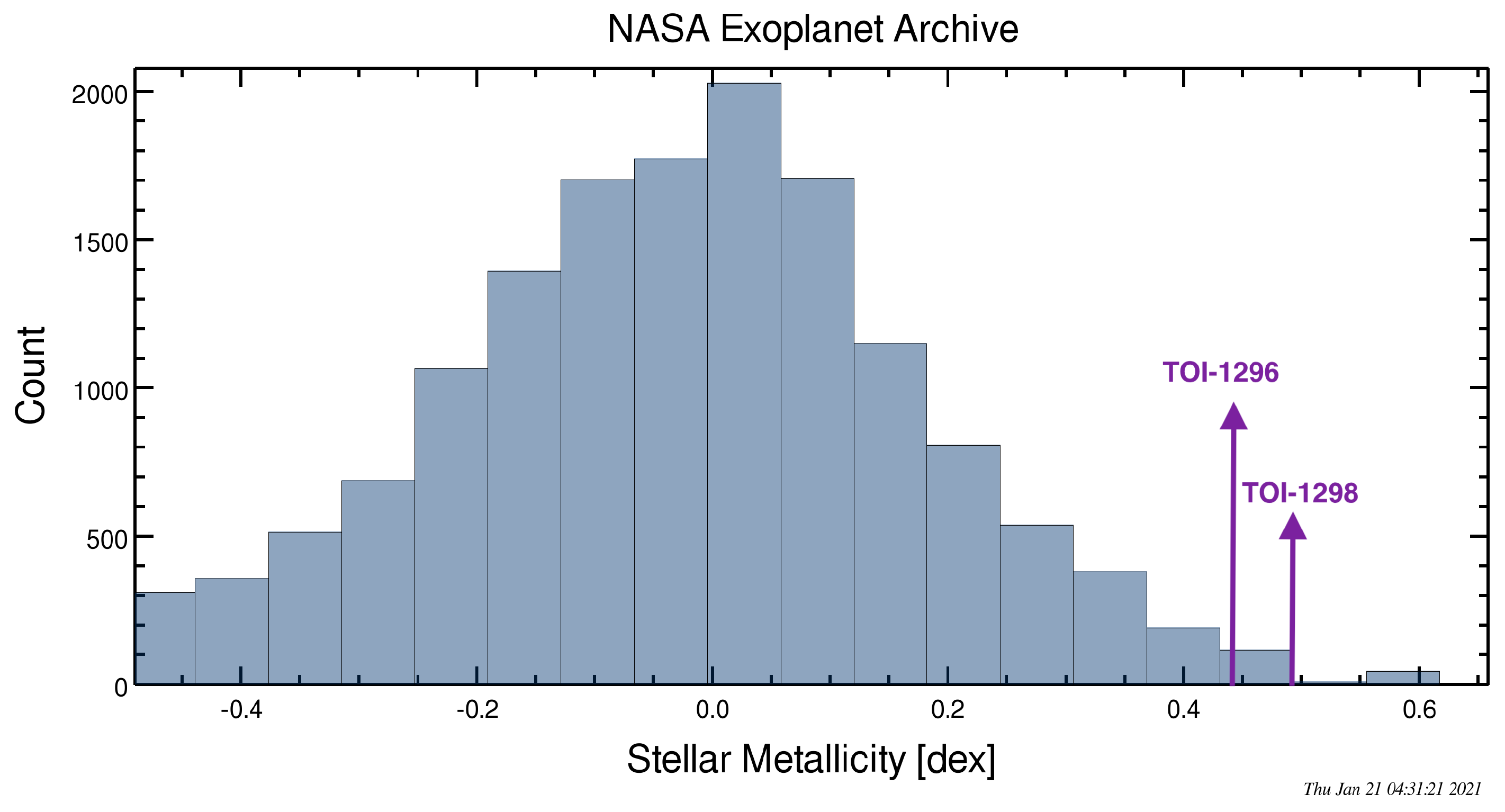}
\caption{Distribution of metallicity in planet host stars from the NASA exoplanet archive$^3$. The extreme properties of TOI-1296 and TOI-1298 are shown with arrows.}
\label{fig:feh}
\end{figure}

The host stars are, moreover, enriched in heavy elements, with metallicities of $+$0.44$\pm$0.04 and $+$0.49$\pm0.03$. Figure \ref{fig:feh} shows the current distribution of stellar metallicity in the population of exoplanet hosts. This distribution shows a larger proportion of solar metallicity and a larger wing towards the low metallicity, as for field stars \citep{holmberg2007}. From Fig. 21 of \citet{holmberg2007}, we can see that less than 0.5\% of stars in the solar vicinity do indeed have a metallicity larger than 0.45. TOI-1296 and TOI-1298 thus stand out in the extreme wing of high metallicity. Stars of such metallic content have the largest odds to host a giant planet in short orbit as shown in early studies \citep{santos2004, fischer2005} and most recently by \citet{petigura2018}, who derived an occurrence rate of about three giant planets per 100 stars in the period range from 1-10 days for host stars as metal-rich as TOI-1296 and TOI-1298 (their Fig. 10a). This rate is about 10 times greater than for stars of solar metallicity, according to this study. The system of TOI-1296 looks similar to K2-97, an inflated hot Jupiter orbiting an evolved, metal-rich star and used as a benchmark for the reinflation process \citep{grunblatt2017}, and also to HATS-41 \citep{bento2018}, Kepler-91 \citep{barclay2015}, and HATS-54 \citep{espinoza2019}. On the other hand, the system TOI-1298 has similar properties as XO-7 \citep{crouzet2020} and WASP-21 \citep{bonomo2017}. With this work, we are adding two transiting systems in the most metal-rich bin, one of them being hosted by an evolved star.

The large-density planet TOI-1298b and its host star is intermediate between Saturn and the most studied transiting system of HD 149026 \citep{sato2005}. HD 149026 is also metal rich, and its high-density planet likely has a massive core. It is also expected for TOI-1298b for its internal core and/or atmospheric composition to be extremely enriched, as shown in \citet{guillot2005}, \citet{fortney2006}, and \citet{baraffe2008}. From irradiated models of planetary internal structure presented in \citet{baraffe2008} with a mass of about 110 Earth masses and an age of 9.5 Gyr, the radius of TOI-1298b would be consistent with a relative content of heavy elements of 15 to 40\% (or 16 to 44 Earth masses). It is impossible to say whether this mass of heavy elements is concentrated in a core or mixed within the envelope due to our current knowledge of giant planet interiors, as described for instance in \citet{debras2019}.

On the other extreme side, TOI-1296b is very fluffy with a density about 4 times smaller, while the stellar host is equivalently metal-rich. Within its mass range, TOI-1296b is one of the most inflated planets, as illustrated in Figure \ref{fig:rprs} (bottom). As it may be due to the reinflation of the planet, it may not automatically imply a low content in heavy elements. The tendency is indeed for high-metallicity stars to have high-metallicity planets \citep{moutou2013}. With reinflation, the new flux income from the stellar luminosity increase may alternatively explain the lower density, despite the presence of heavy elements in the planet atmosphere. The formation, internal structure, and evolution of such diverse planets remains to be precisely described.

\footnotetext{https://exoplanetarchive.ipac.caltech.edu/}

\begin{table}
\renewcommand{\arraystretch}{1.25}
\centering
\caption{SOPHIE RV and errors for TOI-1296.}\label{table.rvs1296}
\begin{tabular}{lcc}
\hline
MJD & RV & $\sigma$\\
    & km/s &  m/s \\
\hline
2459003.49320&24.9370 &0.0090\\
2459016.52095&25.0607&0.0057\\
2459018.48719&24.9841&0.0062\\   
2459019.48308&25.0273&0.0047\\ 
2459020.46828&25.0545&0.0056\\
2459039.43075&25.0404&0.0053\\
2459040.42127&25.0555&0.0036\\
2459056.43518&25.0342&0.0056\\
2459057.43562&25.0116&0.0033\\
2459059.39879&25.0134&0.0052\\   
2459061.42214&24.9853&0.0052\\
2459080.35443&25.0369&0.0053\\
2459139.31406&25.0442&0.0050\\
2459140.36151&24.9850&0.0048\\
2459141.29465&25.0060&0.0052\\
\hline
\end{tabular}
\end{table}

\begin{table}
\renewcommand{\arraystretch}{1.25}
\centering
\caption{SOPHIE RV and errors for TOI-1298.}\label{table.rvs1298}
\begin{tabular}{lcc}
\hline MJD & RV & $\sigma$\\
    & km/s & m/s \\
\hline
2459039.39641  &   -56.0040    &    0.0054\\
2459059.38347  &   -55.9629    &    0.0050\\
2459060.41331  &   -55.9345    &    0.0054\\ 
2459061.39563  &   -55.9808    &    0.0045\\
2459062.38681  &   -56.0070    &    0.0060\\
2459081.36504  &   -55.9724    &    0.0045\\
2459082.38186  &   -55.9578    &    0.0055\\
2459083.34692  &   -55.9602    &    0.0049\\
2459084.39174  &   -56.0133    &    0.0060\\
2459085.32123  &   -56.0255    &    0.0072\\
2459107.32167  &   -56.0114    &    0.0072\\
2459140.27286  &   -55.9745    &    0.00423 \\
2459141.26965  &   -55.9561    &    0.0048\\
2459149.29675  &   -55.9368    &    0.0073 \\
\hline
\end{tabular}
\end{table}

\begin{acknowledgements}
{ 
We are grateful to the anonymous referee who provided very useful criticism and suggestions. 

Funding for the TESS mission is provided by NASA's Science Mission Directorate.

We acknowledge the use of public TESS data from pipelines at the TESS Science Office and at the TESS Science Processing Operations Center.

Resources supporting this work were provided by the NASA High-End Computing (HEC) Program through the NASA Advanced Supercomputing (NAS) Division at Ames Research Center for the production of the SPOC data products.

This paper includes data collected by the TESS mission that are publicly available from the Mikulski Archive for Space Telescopes (MAST).

We acknowledge funding from the French National Research Agency (ANR) under contract number ANR-18-CE31-0019 (SPlaSH).

SH acknowledges CNES funding through the grant 837319. This work was also supported by FCT - Funda\c{c}\~ao para a Ci\^encia e a Tecnologia through national funds and by FEDER through COMPETE2020 - Programa Operacional Competitividade e Internacionaliza\c{c}\~ao by these grants: UID/FIS/04434/2019; UIDB/04434/2020; UIDP/04434/2020; PTDC/FIS-AST/32113/2017 \& POCI-01-0145-FEDER-032113; PTDC/FIS-AST/28953/2017 \& POCI-01-0145-FEDER-028953; PTDC/FIS-AST/28987/2017 \& POCI-01-0145-FEDER-028987. 

PC thanks the LSSTC Data Science Fellowship Program, which is funded by LSSTC, NSF Cybertraining Grant \#1829740, the Brinson Foundation, and the Moore Foundation; her participation in the program has benefited this work.

This work has been carried out within the framework of the NCCR PlanetS supported by the Swiss National Science Foundation. Finally, we thank J. Lissauer for his interest and advice.
}
\end{acknowledgements}

\bibliographystyle{aa}
\bibliography{references}

\begin{appendix} 
\section{TESS apertures}
\begin{figure}
\includegraphics[width=0.9\hsize]{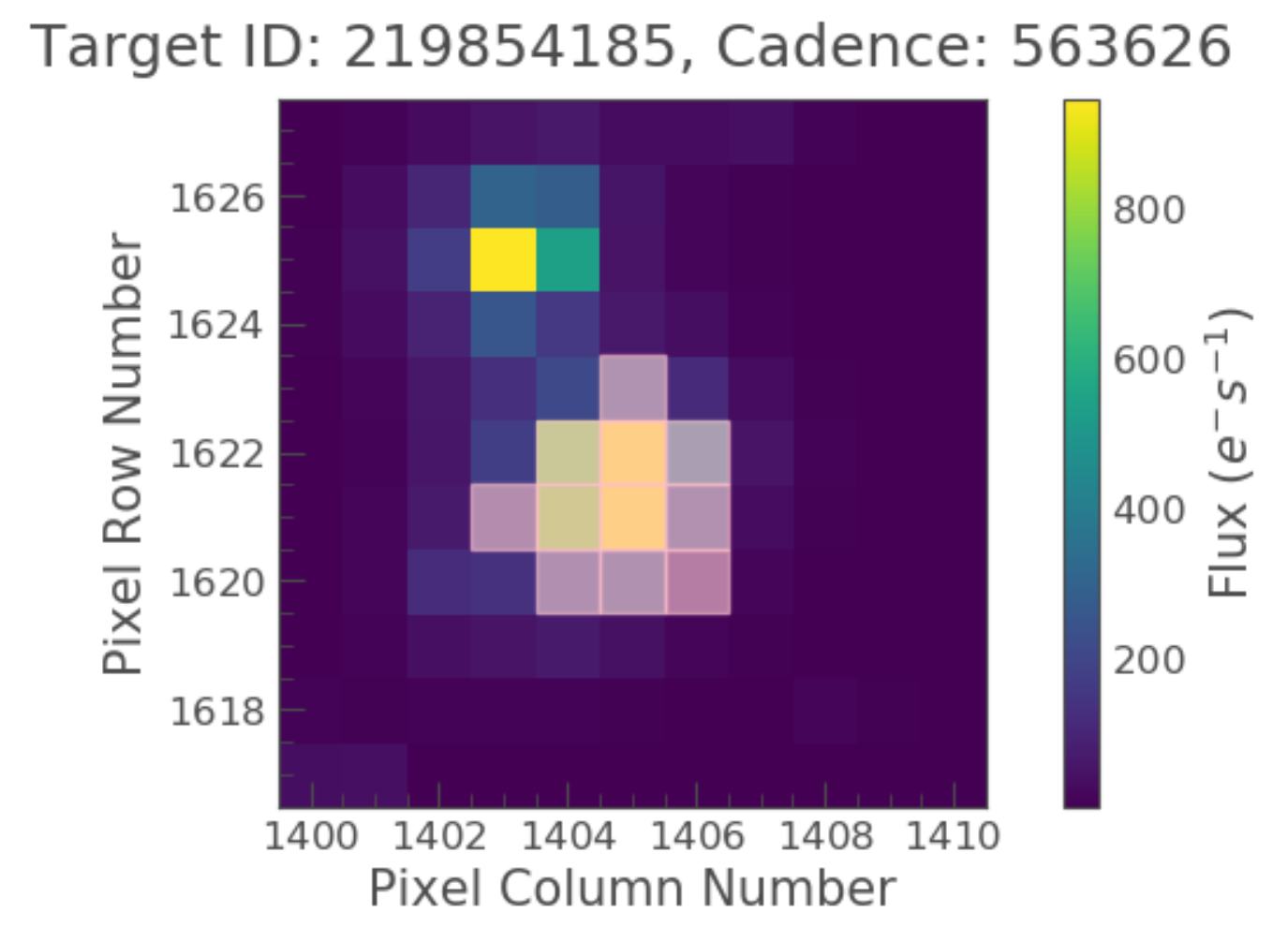}
\includegraphics[width=0.9\hsize]{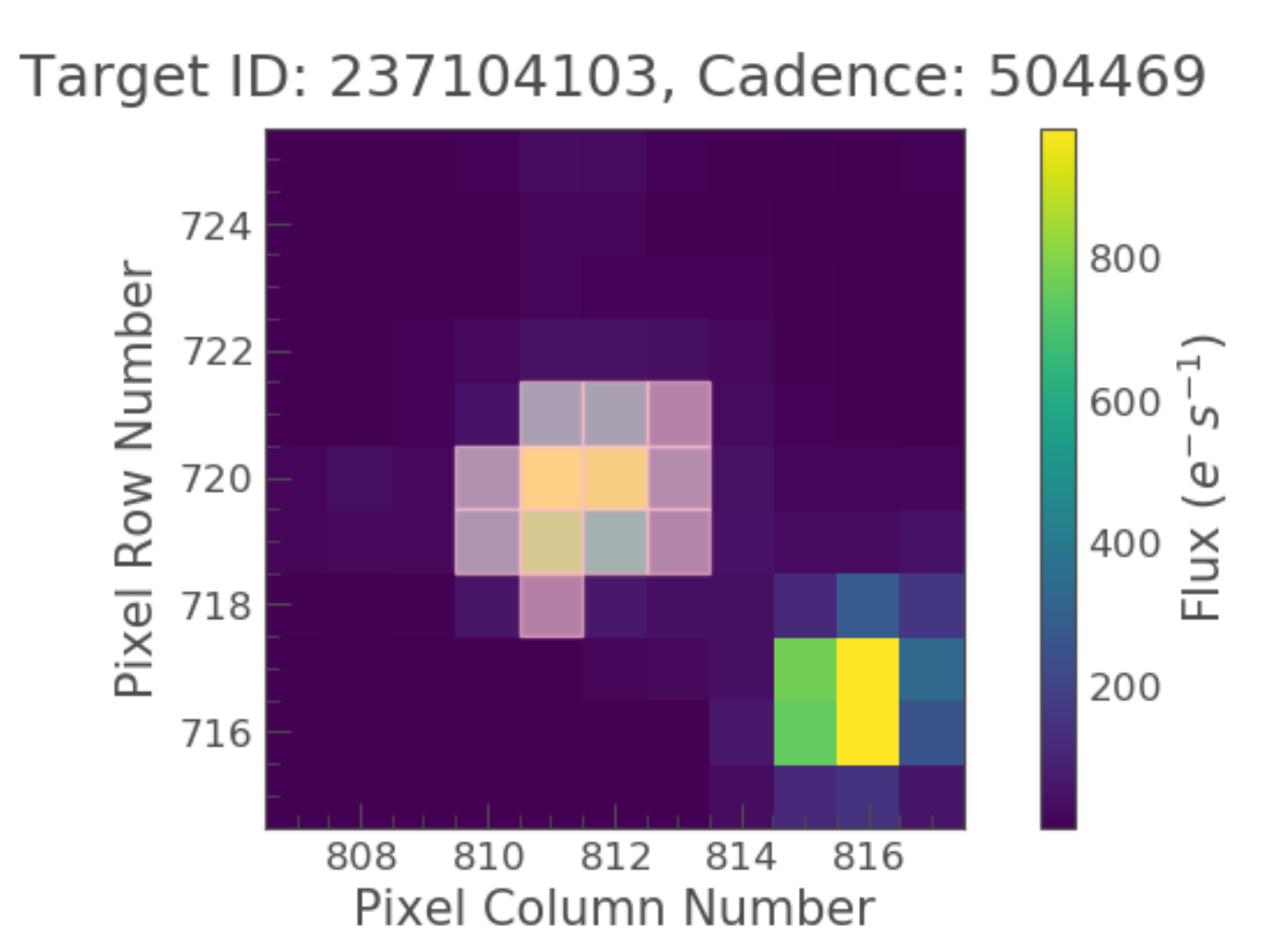}
\caption{TESS imagettes of TOI-1296 (top) and TOI-1298 (bottom) as provided by the \texttt{lightkurve} package. TESS pixels cover a sky aperture of 21~arcsec.}\label{imagettes}
\end{figure}

The TESS full images in the vicinity of the targets TOI-1296 and TOI-1298 are shown in Figure \ref{imagettes}. The pixels coloured in pink are the ones used in the aperture for the PDC-SAP flux. Both stars have a bright neighbour, at a distance large enough so that the contamination of the main target apertures are not affected. A contamination factor of 0.01934 and 0.001912 for TOI-1296 and TOI-1298 were estimated by the SPOC pipeline, respectively. 

\end{appendix}

\end{document}